\documentclass[reprint,prb,superscriptaddress,showpacs]{revtex4-2}%
\usepackage{graphicx}
\usepackage{color}
\usepackage{amsmath}
\usepackage{amsfonts}
\usepackage{lineno}
\usepackage{amssymb}
\usepackage{framed}
\usepackage{array} 
\usepackage{tabularx} 
\usepackage{float}
\providecommand{\U}[1]{\protect\rule{.1in}{.1in}}
\makeatletter
\renewcommand*{\fnum@figure}{{\normalfont\bfseries \figurename~\thefigure}}

\renewcommand*{\@caption@fignum@sep}{\textbf{ : }}
\makeatother
\usepackage{hyperref}
\hypersetup{
    colorlinks=true,
    linkcolor=blue,
    filecolor=magenta,      
    urlcolor=cyan,
}
\begin{document}

\title{ Three-dimensional nature of anomalous Hall conductivity in YMn$_6$Sn$_{6-x}$Ga$_x$, $x\approx 0.55$} 

\author{Hari Bhandari}
\email{hbhandar@nd.edu}
\affiliation{Department of Physics and Astronomy, University of Notre Dame, Notre Dame, IN 46556, USA}
\affiliation{Stravropoulos Center for Complex Quantum Matter, University of Notre Dame, Notre Dame, IN 46556, USA}
\affiliation{Department of Physics and Astronomy, George Mason University, Fairfax, VA 22030, USA}
\author{Zhenhua Ning}
\affiliation{Ames National Laboratory, U.S. Department of Energy, Ames, Iowa 50011, USA}
\author{Po-Hao Chang}
\affiliation{Department of Physics and Astronomy, George Mason University, Fairfax, VA 22030, USA}
\affiliation{Quantum Science and Engineering Center, George Mason University, Fairfax, VA 22030, USA}
\author{Peter E. Siegfried}
\affiliation{Department of Physics and Astronomy, George Mason University, Fairfax, VA 22030, USA}
\affiliation{Quantum Science and Engineering Center, George Mason University, Fairfax, VA 22030, USA}
\author{Resham B. Regmi}
\affiliation{Department of Physics and Astronomy, University of Notre Dame, Notre Dame, IN 46556, USA}
\affiliation{Stravropoulos Center for Complex Quantum Matter, University of Notre Dame, Notre Dame, IN 46556, USA}
\author{Mohamed El. Gazzah}
\affiliation{Department of Physics and Astronomy, University of Notre Dame, Notre Dame, IN 46556, USA}
\affiliation{Stravropoulos Center for Complex Quantum Matter, University of Notre Dame, Notre Dame, IN 46556, USA}
\author{Albert V. Davydov}
\affiliation{Materials Science and Engineering Division, National Institute of Standards and Technology (NIST), Gaithersburg, Maryland 20899, USA}
\author{Allen G. Oliver}
\affiliation{Department of Chemistry and Biochemistry, University of Notre Dame, Notre Dame, IN 46556, USA}
\author{Liqin Ke}
\affiliation{Ames National Laboratory, U.S. Department of Energy, Ames, Iowa 50011, USA}
\author{Igor I. Mazin}
\affiliation{Department of Physics and Astronomy, George Mason University, Fairfax, VA 22030, USA}
\affiliation{Quantum Science and Engineering Center, George Mason University, Fairfax, VA 22030, USA}
\author{Nirmal J. Ghimire}
\email{nghimire@nd.edu}
\affiliation{Department of Physics and Astronomy, University of Notre Dame, Notre Dame, IN 46556, USA}
\affiliation{Stravropoulos Center for Complex Quantum Matter, University of Notre Dame, Notre Dame, IN 46556, USA}

\date{\today}
\begin{abstract}
\textbf{The unique connectivity of kagome lattices gives rise to topological properties, such as flat bands and Dirac cones. When combined with ferromagnetism and a chemical potential near the 2D Dirac points, this structure offers the potential to realize the highly sought-after topological Chern magnetotransport. Recently, there was considerable excitement surrounding this possibility in the ferrimagnetic kagome metal TbMn$_\mathbf{6}$Sn$_\mathbf{6}$. However, density functional theory (DFT) calculations reveal that the 2D Chern gap lies well above the Fermi energy, challenging its relevance in the observed anomalous Hall conductivity. Here, we investigate YMn$_\mathbf{6}$Sn$_\mathbf{5.45}$Ga$_\mathbf{0.55}$, a compound with similar crystallographic, magnetic, and electronic properties to  TbMn$_\mathbf{6}$Sn$_\mathbf{6}$. Our findings show that the intrinsic anomalous Hall conductivity in this material, while comparable in magnitude to that in TbMn$_\mathbf{6}$Sn$_\mathbf{6}$, is fully three-dimensional, thus providing experimental evidence that Hall conductivity in this class of materials does not originate from 2D Chern gaps. Additionally, we confirm that the newly proposed empirical scaling relation for extrinsic Hall conductivity is universally governed by spin fluctuations.}
\end{abstract}

\maketitle
\section{Introduction}\label{sec:1}
\begin{figure}[!ht]
\begin{center}
\includegraphics[width=1\linewidth]{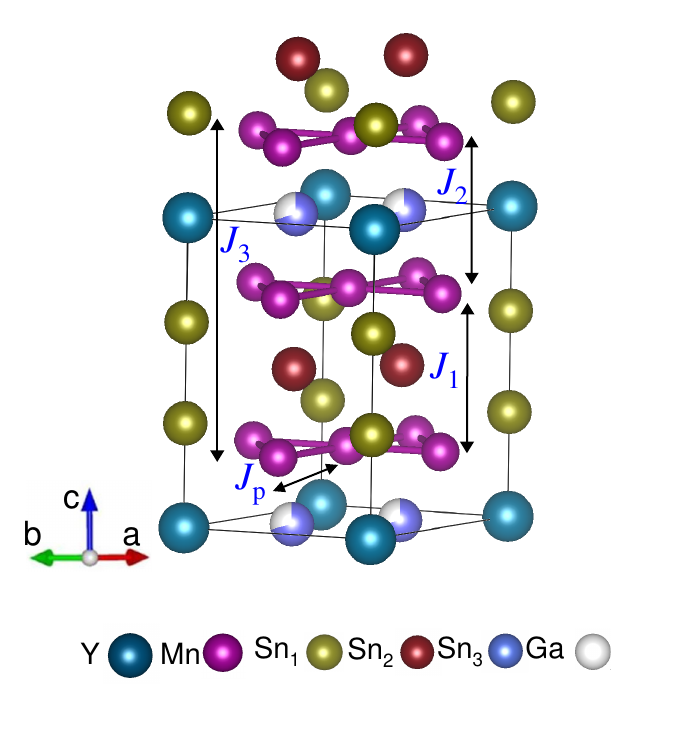}
    \caption{\small\textbf{Crystal structure.} Sketch of the crystal structure of YMn$_6$Sn$_{5.45}$Ga$_{0.55}$. $J_i$s ($i=1,2,3)$ represent the exchange constants and $J_p$ represents the in-plane exchange constant. }
    \label{structure}
    \end{center}
\end{figure}

Kagome lattice magnets have attracted significant interest in condensed matter physics due to their high frustration in the case of antiferromagnetic interactions. Over the past decade, this interest has grown, as it has been shown that even unfrustrated ferromagnetic (or nonmagnetic) kagome planes can exhibit nontrivial electronic features, such as flat bands and Dirac cones \cite{zhang2011quantum,ghimire2020topology,kang2020dirac,ye2018massive,yin2019negative,kuroda2017evidence,superconductivity}. Recently, particular attention has been directed toward the so-called 166 family, the $R$Mn$_6$Sn$_6$ compounds, where $R$ represents a rare-earth element \cite{bolens2019topological,asaba2020anomalous,bhandari2024magnetism,pokharel2021electronic,jones2022origin,pokharel2022highly,arachchige2022charge,riberolles2023orbital,lee2022interplay}. In these compounds, Mn atoms form kagome planes, and the crystal structure provides a diverse material space for manipulating both electronic and magnetic properties (Fig. \ref{structure}). Compounds with non-magnetic $R$ atoms are simpler because magnetism arises solely from the Mn sublattice; however, they are also more complex due to frustrated interplanar magnetic interactions. As shown in Fig. \ref{structure}, although all Mn planes are crystallographically equivalent, the exchange interactions are not; there are two distinct exchange paths: one ($J_1$) across the Sn layer and the other ($J_2$) across the $R$Sn layer. In  YMn$_6$Sn$_6$ (Y166), a $R$166 compound with non-magnetic $R$ atom,  $J_2 > 0$ (i.e., antiferromagnetic), while the dominant interaction is $J_1 < 0$. As in most metals, the exchange coupling decays relatively slowly with distance (roughly as 1/$d^3$), so the minimal model Hamiltonian includes $J_1$, $J_2$, and $J_3$. Notably, this Hamiltonian is frustrated if $J_2J_3 < 0$, resulting in intriguing spin-spiral orders that exhibit phenomena such as the topological Hall effect \cite{ghimire2020competing,wang2021field} and Lifshitz transitions \cite{siegfried2022magnetization}.

Replacing Y with a magnetic ion that strongly couples to the neighboring Mn planes introduces an additional indirect ferromagnetic coupling between these planes, which can overcome the direct antiferromagnetic $J_2$ and remove magnetic frustration. This 
occurs in Tb166 \cite{riberolles2023orbital,yin2020quantum,riberolles2022low,mielke2022low,li2023discovery,xu2022topological,gao2021anomalous} and is now well understood \cite{lee2022interplay,jones2022origin}. Additionally, due to well-understood reasons \cite{lee2022interplay}, Tb has an easy axis anisotropy along the crystallographic $c$ direction, making the material a collinear easy-axis ferrimagnet at low temperatures (in contrast to Y166, which is an easy-plane spiral antiferromagnet). Because of these specific characteristics, Tb166 recently garnered significant interest for various different properties \cite{yin2020quantum, jones2022origin,mielke2022low,xu2022topological,li2023discovery,Wenzel2022Effect} including a putative two-dimensional (2D) Chern gap around 130 meV above the Fermi energy, as inferred from scanning tunneling measurements \cite{yin2020quantum}. One of the consequences of a 2D Chern gap, when close to the Fermi energy, is 
large anomalous Hall effect. The intrinsic contribution to the anomalous Hall effect extracted from the scaling of the anomalous Hall resistivity matches the value expected from the Chern gap around 130 meV above the Fermi energy \cite{yin2020quantum}. However, density functional calculations \cite{jones2022origin,zhang2022exchange,lee2022interplay} consistently have shown that the 2D-Chern gap in Tb166 lies about 700 meV above the Fermi energy and does not contribute to the anomalous Hall conductivity (AHC). Instead, this AHC arises from several different regions in the Brillouin zone in ferrimagnetic Tb166 \cite{jones2022origin}. Additionally, an empirical scaling relation in the latter study suggested a potential contribution of spin fluctuations to the extrinsic anomalous Hall conductivity, an effect not previously investigated.

Regarding the latter, the standard method based on the theory by Cr\'epieux and Bruno \cite{PhysRevB.64.014416} for extracting the intrinsic (related to the Berry phases of conducting electron) and extrinsic (due to electron scattering) contributions to the AHC is to fit the temperature dependence of the AHC to a simple scaling relation, 
\begin{equation}
\sigma_{xy} = a\sigma^2_{xx}+ c \label{scaling0} 
\end{equation}
where $c=\sigma^A_{xy}$ represents the intrinsic AHC, while the first term arises from defect scattering. This scaling relation has been widely used 
to extract the intrinsic AHC in various materials \cite{Tian2009,ye2018massive,yin2020quantum}. However, Cr\'epieux and Bruno's derivation did not account for scattering from thermally excited spin fluctuations, and currently a theoretical analysis of the effect of spin fluctuations on AHC is lacking.

Analyzing the Tb166 data, we found \cite{jones2022origin} that Eq. \ref{scaling0} poorly describes $ \sigma_{xy}$ at higher temperatures, where scattering from spin fluctuations becomes increasingly important. At the same time, we found that the most fluctuating species in Tb166 is Tb, which fluctuates much  stronger than Mn. We also found that the empirical formula,
\begin{equation}
\sigma_{xy} = a\sigma^2_{xx}+ d/\sigma_{xx}+c, \label{scaling1}%
\end{equation}
fits the experimental data exceptionally well (note that the additional contribution becomes significant when $\sigma_{xx}$ is small, i.e., at higher temperatures). We tentatively attributed this term to spin fluctuations, primarily from Tb. 

A reliable protocol for extracting AHC from the experiment is crucial. The existing methods \cite{PhysRevB.64.014416,Tian2009,PhysRevB.96.144426} overlook spin fluctuations, and therefore rely on low-temperature data. An equation that accurately describes the AHC across the entire temperature range is of significant practical importance.

In this study, we address two key aspects concerning Tb166: first, whether the intrinsic contribution to the AHC is linked to the 2D Chern gap, and second, whether the component in the AHC scaling relation, attributed to spin fluctuations, can be observed in another $R$166 compound that does not contain Tb. This is particularly relevant as Tb was thought to be crucial for
both the Chern-gap induced AHE \cite{yin2020quantum} and the enhancement of spin fluctuations \cite{jones2022origin}. 

An ideal compound to investigate these properties would be a $R$166 compound with a non-magnetic $R$ atom and a soft, and relatively isotropic ferromagnet. Such a compound would allow the access to the saturated ferromagnetic state in a standard laboratory setting, enabling measurements of anomalous Hall resistivity over a broad temperature range and in both in-plane and out-of-plane directions. We found that YMn$_6$Sn$_{6-x}$Ga$_x$, $0.30\leq x \leq 0.61$ meet these criteria \cite{xu2021magnetic} and selected one particular composition YMn$_6$Sn$_{6.45}$Ga$_{0.55}$ for the study. Our results show that the intrinsic AHC in this compound is comparable to that in Tb166 for the out-of-plane magnetic field ($B$), where the Chern gap is expected to contribute to the AHC. 
At the same time, we observed a similar AHC in the in-plane $B$, where the Chern gap contribution is not expected, confirming the 3D nature of the AHC. Furthermore, we verified the new AHC scaling introduced in Ref. \cite{jones2022origin} and confirmed its spin-fluctuational origin.

\section{Methods}\label{sec:2}

Single crystals of YMn$_6$Sn$_{5.45}$Ga$_{0.55}$ were grown by using Sn as a flux by the molten flux method. Y pieces (Alfa Aesar 99.9\%), Mn pieces (Alfa Aesar 99.95\%), Sn shots (Alfa Aesar 99.999\%) and Ga pieces (Alfa Aesar 99.9999\%)  were added into a 2-ml aluminum oxide crucible in molar ratio of 1:6:18:2. The crucible was then sealed in a fused silica ampule under vacuum. The sealed ampule was heated to 1150 $^{\circ}$C over 10 hours, homogenized at 1150 $^{\circ}$C for 24 hours, and then cooled down to 600 $^{\circ}$C with a rate of 5 $^{\circ}$C/h. Once the furnace reached 600 $^{\circ}$C, the molten flux was separated from the crystals by using a centrifuge. Upon opening the crucible, nice hexagonal-looking crystals up to 20 mg were obtained. 

The crystal structure and the atomic composition were verified from a single crystal X-ray diffraction experiment. An arbitrary sphere of data was collected on a silver block-like crystal, having appropriate dimensions of $0.073\times0.068\times0.026$, on a Bruker D8 diffractometer equipped with a Bruker APEX-II detector using a combination of $\omega$- and $\phi$-scans of 0.5 degree \cite{apex4}. Data were corrected for absorption and polarization effects and analyzed for space group determination \cite{krause2015comparison}. The structure was solved by dual-space methods and expanded routinely \cite{sheldrick2015shelxt}. The model was refined by full-matrix least-squares analysis of F$^2$ against all reflections \cite{sheldrick2015crystal}. All non-hydrogen atoms were refined with anisotropic atomic displacement parameters. Unless otherwise noted, hydrogen atoms were included in calculated positions. Atomic displacement parameters for the hydrogen were tied to the equivalent isotropic displacement parameter of the atom to which they are bonded ( $U_{iso}(H) =1.5 U_{eq}(C)$ for methyl, $1.2 U_{eq}(C)$ for all others.  

DC magnetization, resistivity, and magnetoresistance measurements were performed in a Quantum Design Dynacool Physical Property Measurement System (PPMS) with a 9 T magnet. ACMS II  option was used in the same PPMS for DC magnetization measurements. Single crystals of YMn$_6$Sn$_{5.45}$Ga$_{0.55}$ were polished to adequate dimensions for electrical transport measurements. Crystals were oriented with the  [001] and [100] directions parallel to the applied field for the c-axis and ab-plane measurements. Resistivity and Hall measurements were done using the 4-probe method. Pt wires of 25 $\mu$m were used for electrical contacts with contact resistances $<$ 30 $\Omega$. Contacts were affixed with Epotek H20E silver epoxy. An electric current of 4 mA was used for the electrical transport measurements. Contact misalignment in the Hall resistivity measurement was corrected by anti-symmetrizing the measured data in positive and negative magnetic fields. 

The first-principles calculations were performed using Vienna ab initio
Simulation Package (VASP) \cite{Kresse1996} within projector augmented
wave (PAW) method \cite{blochl1994,Kresse1999} The Perdew-Burke-Enzerhof
(PBE) \cite{PBE} generalized gradient approximation was employed
to describe exchange-correlation effects. The on-site Coulomb interactions
are taken into account using LDA+U \cite{Anisimov1991} to improve
the description of the interactions between localized $3d$-electrons
of Mn and an effective $U_{eff}=U-J=0$, $0.6$ and $2$ are
considered. 

The experimental values were used for the lattice parameters and kept
fixed for all the calculations, including the geometry optimization,
where only internal coordinates were relaxed. To properly determine
the structure, we performed geometry optimization for $2d$, $2e$, and $2c$
three different Ga substitution sites. The 2c site, consistent with
the experimental analysis, gives the lowest energy which is then considered
as the magnetic ground state.

\section{Results and Discussion}\label{sec:3}

\subsection{Crystallography}\label{sec:3a}

The crystal structure and atomic composition of YMn$_6$Sn$_{5.45}$Ga$_{0.55}$ (Y166-Ga) were determined using single crystal X-ray diffraction. Similar to the parent compound YMn$_6$Sn$_6$ (Y166), Y166-Ga adopts a hexagonal $P6/mmm$ structure $( a = b = 5.4784$\ \AA, $c = 8.925$\ \AA), consisting of kagome planes [Mn$_3$Sn] separated by two inequivalent layers Sn$_2$ and Sn$_3$YGa, as illustrated in Fig. \ref{structure}.

The structure exhibits Sn-site doping with Ga, specifically at the Sn3 site (Wyckoff position 2c). This was confirmed through modeling efforts for partial Ga occupancy at other Sn sites, which either resulted in poorer fits to the data or nonsensical Ga occupancy values. Refinement analysis determined the Sn:Ga site occupancy to be 0.725:0.175, corresponding to the full chemical formula noted above, i.e., $\sim$ 0.55 Ga atoms per unit cell. Detailed crystallographic parameters from the single crystal X-ray diffraction experiment are summarized in Table \ref{T1}.

Our DFT calculations, presented in Sec. \ref{theory}, also reveal a significant energy advantage for Ga substitution at the Sn3 site in the [Sn$_3$Y] layer, rather than in the [Mn$_3$Sn] or [Sn$_2$] layers. YMn$_6$Sn$_{5.45}$Ga$_{0.55}$ can thus be viewed as a moderately hole-doped ($\approx 0.09$ h/Mn) derivative of the parent compound Y166.

\begin{table}[h]
\caption{Crystallographic data, atomic coordinates and equivalent displacement parameters for YMn$_6$Sn$_{5.45}$Ga$_{0.55}$.
}\label{T1} 
\renewcommand{\arraystretch}{1}
\small 
\begin{tabular}
 [c]{l@{\hspace{0.15cm}}l@{\hspace{0.15cm}}l@{\hspace{0.15cm}}l@{\hspace{0.15cm}}l@{\hspace{0.15cm}}l}\hline   
Crystal system & Hexagonal \\
Space group & $P6$/mmm  \\
Temperature (K) & 120(2)  \\
Wavelength (\AA) & 0.71073  \\
Z formula units & 1      \\

2 $\theta$ $_{min}$ & 4.564$^\circ$  \\
2 $\theta$ $_{max}$ & 61.206$^\circ$           \\ 
Formula weight (g/mol) & 1103.76  \\
a,b (\AA) & 5.4784(16)\\
c (\AA)   & 8.925(4)      \\
Volume (\AA)$^3$ &231.97(17)             \\
Density (calculated) (g/cm$^3$) & 7.901                \\
$\mu$ (Mo$K_{\alpha}$) mm$^{-1}$ & 29.896              \\
Goodness-of-fit on $F^2$ & 1.186  \\
Final R indices [I$\ge$2$\sigma$ (I)] &R$_1$ = 0.0172  \\
 &wR$_2$ = 0.0392\\
R indices (all data)  &R$_1$ = 0.0183 \\
 & wR$_2$ = 0.0397\\
Largest diff. peak \& hole, e (\AA$^{-3}$) & 1.128 and -1.341 \\
\end{tabular}

\par%

\begin{tabular} 
[c]{c@{\hspace{0.2cm}}c@{\hspace{0.2cm}}c@{\hspace{0.2cm}}c@{\hspace{0.2cm}}c@{\hspace{0.2cm}}c}\hline

 Atom &Wyck & x & y & z & Ueq ${\AA}^2$     \\ \hline
Y   &   1a  &   1.00000   & 1.00000      &   0.00000 & 0.007(1) \\
Sn(1)  &  2e  &  1.00000   &  1.00000   &   0.33674(8) & 0.004(1) \\
Sn(2)   &  2d   &   0.33333   & 0.66667     &   0.50000   & 0.006(1) \\
Sn(3)  &  2c   & 0.66667    &  0.33333  &  0.000000    & 0.005(1)\\ 
Ga(1)  &  2c   & 0.6667     &  0.33333  &  0.000000    & 0.005(1)\\ 
Mn(1)  &  6i   & 0.50000     & 0.50000    &  0.24504(9)   & 0.005(1)\\ 

 \hline

\end{tabular}
\end{table}

\subsection{Magnetic Properties}\label{sec:3b}

\begin{figure*}[ht!]
\begin{center}
\includegraphics[width=1\linewidth]{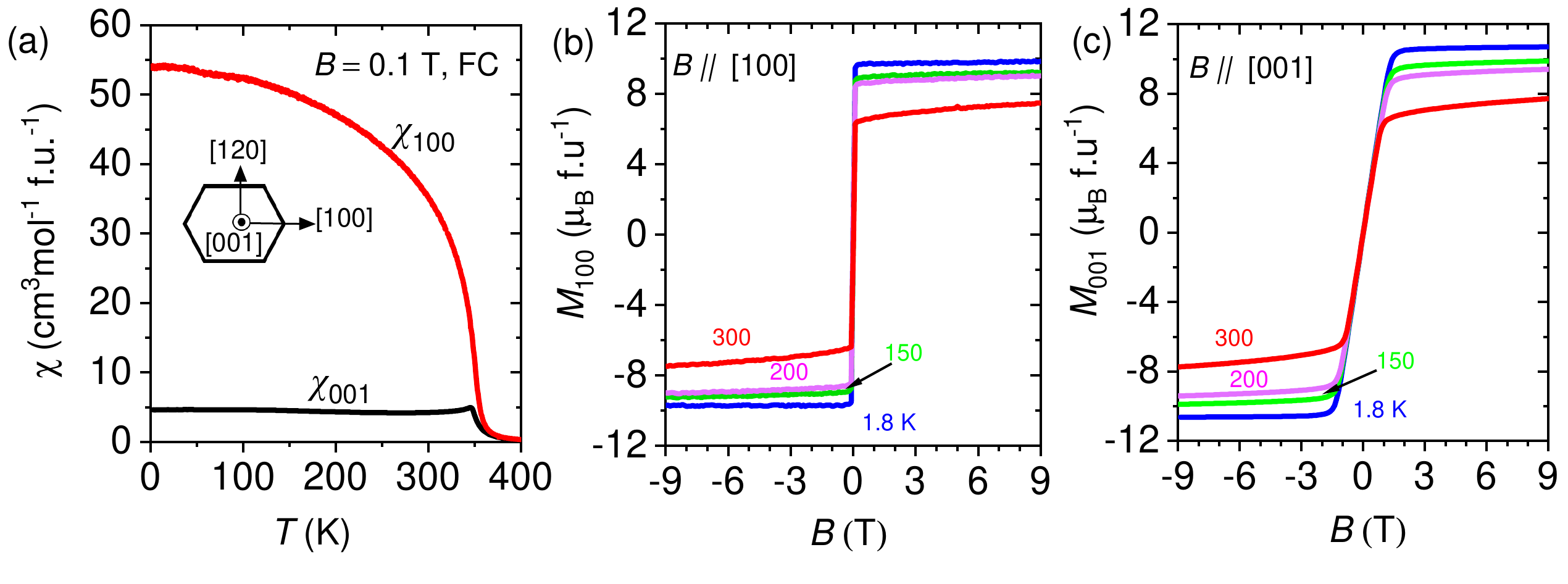}
    \caption{\small\textbf{Magnetic properties of YMn$_\mathbf{6}$Sn$_\mathbf{5.45}$Ga$_\mathbf{0.55}$.} (a) Magnetic susceptibility measured under an external magnetic field $B$ of 0.1 T along the [100] and [001] directions using the field-cooled protocol. (b-c) Magnetization as a function of magnetic field $B||$ [100] (b) and $B||$ [001] (c) at selected temperatures ranging from 1.8 to 300 K.}
    \label{susc}
    \end{center}
\end{figure*}

The temperature dependence of magnetic susceptibility ($\chi = M/B$) measured with a magnetic field $B$ of 0.1 T parallel to [100] ($\chi_{100}$) and along [001] ($\chi_{001}$) is shown in Fig \ref{susc} (a). These susceptibility data indicate that Y166-Ga undergoes a paramagnetic-to-ferromagnetic ordering below 350 K, consistent with previous reports \cite{xu2021magnetic, zhang2001magnetism}. Additionally, the easy-plane behavior is evident from the significantly larger $\chi_{100}$ compared to $\chi_{001}$ below the transition temperature ($T_c$), with an anisotropy ratio of  $\chi_{100}/\chi_{001} = 5$ just below $T_c$. It is to be noted that parent compound Y166 orders with a commensurate antiferromagnetic helical structure below 345 K and exhibits an incommensurate double helical structure (DH) upon further cooling \cite{ghimire2020competing,siegfried2022magnetization,dally2021chiral}, while YMn$_6$Sn$_{1-x}$Ga$_{x}$ compounds show a ferromagnetic transition for doping concentrations $x>0.30$ \cite{xu2021magnetic}.

Figures \ref{susc} (b) and (c) show the isothermal magnetization curves of Y166-Ga at some representative temperatures for $B$ $||$ [100] ($M_{100}$) and [001] ($M_{001}$), respectively. In the entire temperature range measured $M_{100}$ saturates below 0.5 T with a negligible hysteresis, while $M_{001}$ saturates at slightly larger $B$. At 1.8 K $M_{001}$ saturates at 2.2 T, which decreases with increasing temperature (1.7 T at 300 K), also with negligibly small hysteresis. At 1.8 K, saturation magnetization ($M_{sat}$) along [100] is 9.85 $\mu$$_B$/f.u. while it is 10.7 $\mu$$_B$/f.u along [001]. In either direction, $M_{sat}$ gradually decreases with the increase in temperature, which attains a value of 7.5 $\mu$$_B$/f.u. along [100]  and 7.7 $\mu$$_B$/f.u along [001]. The ratio of the saturated magnetization at 1.8 to 300 K  ($M_{sat,1.8K}$/ $M_{sat,300K}$) is 1.39 along the [001] direction and 1.31 along the [100] direction. For Tb166, the Mn moment shows the ratio of  $\sim$ 1.05 and the Tb moment exhibits the ratio of around 1.66 \cite{Idrissi1991, jones2022origin}. This suggests that the Mn moments in Y166-Ga experience more fluctuation between 1.8 and 300 K than in Tb166, where the Tb moments are the most fluctuating ones \cite{jones2022origin}. This observation is consistent with the expectation, as the Curie temperature of Tb166 is about 70 K higher than that of Y166-Ga.

\subsection{First Principles Calculations}\label{theory}

\begin{table}
\renewcommand*{\arraystretch}{1.4}

\begin{tabular}{ccccc}
\hline 
 & \multicolumn{3}{c}{$x=1$} & $x=0$ (Y166)\tabularnewline
\hline 
U & 0 & 0.6 & 2 & 0.6 \tabularnewline
\hline 
\hline 
$J_{1}$ & -14.5 & -11.9 & -17.3 &   -12.86 \tabularnewline
$J_{2}$ & -21.7 & -10.6 & -21.9 &    4.66  \tabularnewline
$J_{3}$ &   1.8 &   1.5 &  -4.0 &   -2.20  \tabularnewline
\hline 
\end{tabular} 

\caption{\label{tab:jex} Calculated exchange couplings $J_1$, $J_2$ and $J_3$ in unit of meV. The data for Y166 are taken from Ref. \cite{ghimire2020competing}.}
\end{table}

To understand the doping-induced phase transition, we adopt the $J_1-J_3$ effective model proposed
in Ref. \cite{ROSENFELD20081898}. The spin Hamiltonian is expressed as follows
\begin{equation}
H  =\sum_{\left\langle ij\right\rangle _{1}}J_{1}S_{i}S_{j}
   +\sum_{\left\langle ij\right\rangle _{2}}J_{2}S_{i}S_{j}
   +\sum_{\left\langle ij\right\rangle _{3}}J_{3}S_{i}S_{j}
\label{eq:spinham}
\end{equation}
where $J_{i}$ ($i=1-3)$ are the exchange interaction parameters as
indicated in Fig. \ref{structure}. The parameters were extracted using the
least squared fitting of our DFT calculations into Eq. \ref{eq:spinham}.

According to a thorough DFT-based analysis of the phase diagram for
parent Y166 discussed in our earlier work \cite{ghimire2020competing}, we
observed that $U_{eff}=0.6$ best reproduce the magnetic states
for parent Y166 observed in the experiments. Therefore, in our study,
we considered the same $U$ parameter with one additional larger $U=2$
for comparison.

The results along with those of Y166, taken from Ref. \cite{ghimire2020competing}, are summarized in TABLE \ref{tab:jex}. The fitting is achieved with excellent quality in all three cases, as all the relative energy differences between different magnetic states can be consistently and accurately reproduced by the model, especially in the case of $U=0.6$ where the average error is less than 1 meV. This suggests that the minimal model adopted here is appropriate and reliable. 

In TABLE \ref{tab:jex}, one can see that while $J_{1}$ and $J_{2}$ consistently
favor ferromagnetic alignment regardless of different U values, the Hubbard
U correction tends to stabilize the FM states further as $J_{3}$ shifts
from positive to negative as U increases. According to the
analytically determined phase diagram Ref. \cite{ROSENFELD20081898} the spin
model for all three U values yields the same correct FM ground state.

To assess the doping effect, we compare the results of the two
compounds for the same $U=0.6$ (i.e. columns 3 and 5). In Y166, $J_{1}$
dominates and has the same (opposite) sign as $J_{3}$ ($J_{2}$).
This arrangement leads to a competition between $J_{2}$ and $J_{3}$ which
results in the formation of a helical magnetic state \cite{ghimire2020competing,ROSENFELD20081898}.
However, with Ga-doping, $J_{2}$ becomes ferromagnetic and now comparable
to $J_{1}$ in strength. Although $J_{3}$ becomes antiferromagnetic,
it is too weak to induce frustration.

This qualitative shift aligns with expectations, considering the direct alteration
of exchange pathways for $J_{2}$ and $J_{3}$ induced by the presence
of doped-Ga. As a consequence of these changes, the frustration that
was initially developed in the pure Y166 to promote the helical magnetic
state is effectively mitigated. The system undergoes a transition,
and the magnetic state collapses into a ferromagnetic (FM) order.

It is interesting to note that while all configurations predict the
same correct ferromagnetic ground state, only $U=2$ gives the correct
easy-plane anisotropy and $U=0$ and $0.6$ give very small easy-axis
anisotropy.
\begin{figure*}[!ht]
\begin{center}
\includegraphics[width=1\linewidth]{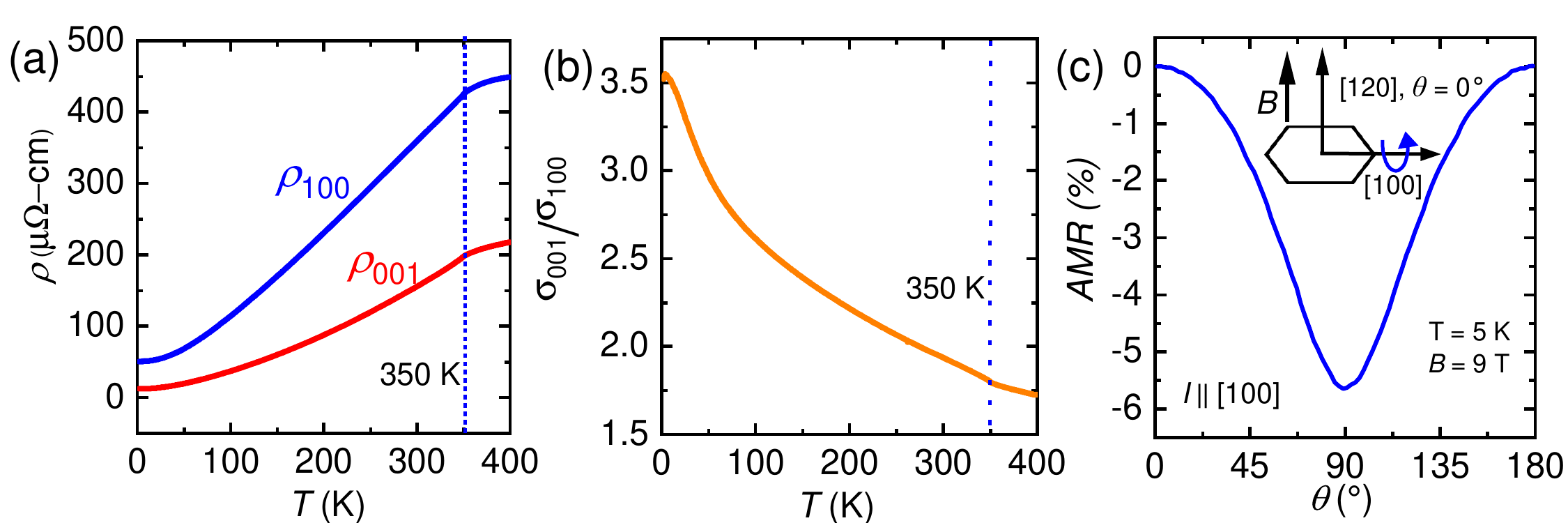}
    \caption{\small\textbf{Electrical transport properties of YMn$_\mathbf{6}$Sn$_\mathbf{5.45}$Ga$_\mathbf{0.55}$.} (a) Electrical resistivity as a function of temperature for current applied along the [100] and [001] directions. (b) Ratio of the conductivity along the [001] direction to that along [100]. (c) Angular magnetoresistance (AMR) as a function of angle $\theta$, measured under a 9 T magnetic field, where the current ($I$) is along the [100] direction and  $\theta$ is the angle between the magnetic field and [120] direction, as illustrated by the sketch in panel (c).}
    \label{res}
    \end{center}
\end{figure*}

\subsection{Electrical Resistivity and Conductivity }

The temperature dependence of electrical resistivity of Y166-Ga, measured with the electric current applied along the [100] direction ($\rho_{100}$, blue curve) and the [001] direction ($\rho_{001}$, red curve) over the  temperature range 1.8-400 K,  is shown in Fig. \ref{res}(a). The resistivity decreases as temperature decreases, indicating the metallic behavior of the sample. Residual resistivity ratio (RRR), calculated as
$\rho(400 K)/\rho(2 K)$,  is 9 for $I||[100]$ and is 18 for $I||[001]$. These values are smaller than those in Y166 \cite{ghimire2020competing,siegfried2022magnetization}, likely due to disorder induced by doping. Both $\rho_{100}$ and $\rho_{001}$ exhibit a kink at 350 K, indicative of the onset of a ferromagnetic transition, as observed in the susceptibility measurements [Fig. \ref{susc}(a)]. Across the entire temperature range, $\rho_{100}$ is greater than $\rho_{001}$. The conductivity anisotropy, $\sigma_{[001]}/\sigma_{[100]}$, is plotted in Fig. \ref{res}(b), showing a value $>$ 2, which suggests that the electronic transport in Y166-Ga is three dimensional. This behavior differs from that of the parent compound Y166, where in-plane conductivity is greater than the out-of-plane conductivity \cite{siegfried2022magnetization}). The enhanced $c$-axis conductivity observed in Y166-Ga is similar to that found in Ge doped YMn$_6$Sn$_6$ \cite{bhandari2024magnetism}. 

The anisotropic transport behavior was further investigated by measuring the angular-dependent magnetoresistance (AMR). In this measurement, an electric current $I$ was applied along the [100] direction, while the sample was rotated around the magnetic field within the crystallographic $bc$-plane, with $I\bot B$ held constant so that only transverse MR was measured. In this configuration, at $\theta = 0^{\circ}(90^{\circ})$, $B||[120]([001])$. Since the largest magnetic saturation field is below 2.5 T (see Fig. 2), the 9T magnetic field aligns the magnetic moment $M$ perpendicular to $I$ at all times.

\begin{figure}[!ht]
\begin{center}
\includegraphics[width=1\linewidth]{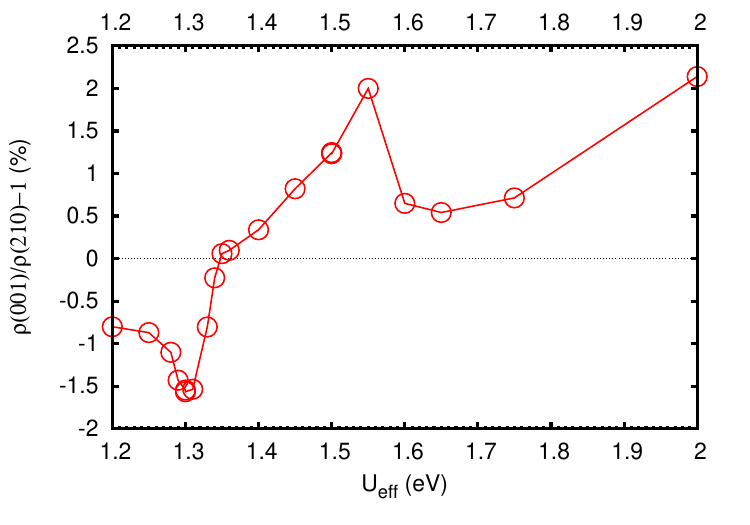}
\caption{\small\textbf{Calculated angular magnetoresistance.} Angular magnetoresistance between the full polarization along the crystallographic [001] and [120] axes, as a function of the effective Hubbard interaction, calculated for YMn$_6$Sn$_{5}$Ga.}
\label{MA}
\end{center}
\end{figure}

The AMR, defined as [\{$\rho_{xx}(\theta)-\rho_{xx}(\theta=0)\}/\rho_{xx}(\theta=0)]\times100$, measured at 5 K is shown in fig. \ref{res}(c), with maximum value of  $-5.6\%$, indicating a substantial effect. To gain further insight into this large AMR, we compared the AMR with the ab initio calculated AMR using the GGA+U method. For this purpose, we used the all-electron WIEN2k package \cite{WIEN}, varying $U_{eff}=U-J$ from 1.2 to 2 eV. Assuming an isotropic transport scattering rate, the longitudinal conductivity 
$\sigma_{xx}\propto \omega_{pl}^2\propto 
\sum_\mathbf{k}{\delta(E-
E_\mathbf{k})v_{\mathbf{k}x}^2}$, where $v$ is 
the Fermi velocity. We calculated this quantity using the $19\times19\times11$ zone-centered k-point mesh and tetrahedron numerical integration, and an otherwise default setup. We found (Fig. \ref{MA}) that AMR is very sensitive to correlation effects. The best agreement with the experiment occurs at $U_{eff}=1.3$ eV, yielding a calculated value of $1.6$\%, which is over three times smaller than the experimental result but remains the same order of magnitude. This discrepancy may indicate the presence of an anisotropic scattering rate, or an underestimation of spin-orbit coupling (SOC) in the calculation, potentially related to the known underestimation of the   AHC in TbMn$_6$Sn$_6$ \cite{jones2022origin,Binghai}.

\subsection{Anomalous Hall Effect }

\begin{figure*}[ht!]
\begin{center}
\includegraphics[width=0.9\linewidth]{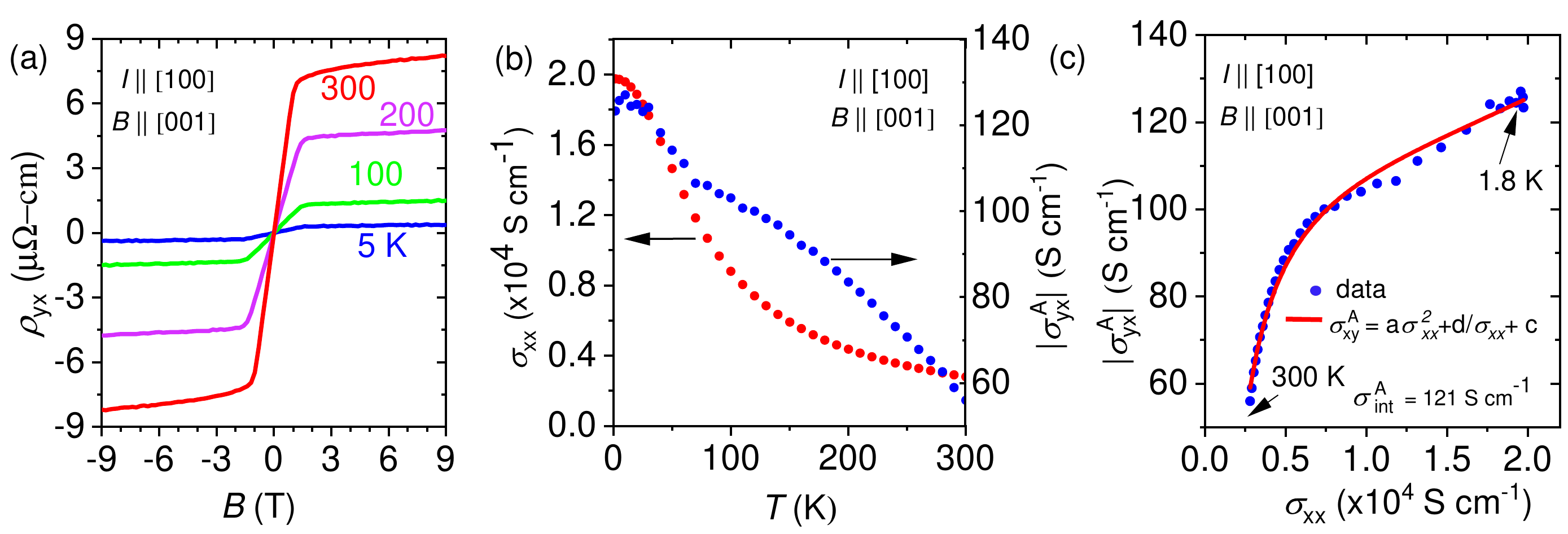}
    \caption{\small\textbf{Anomalous Hall effect of YMn$_\mathbf{6}$Sn$_\mathbf{5.45}$Ga$_\mathbf{0.55}$ measured with $\mathbf{B}~\boldsymbol{||}$ [001] and $\mathbf{I}~\boldsymbol{||}$ [100].} (a) Anomalous Hall resistivity $\rho_{yx}$ as a function of magnetic field at selected temperatures. (b) Longitudinal conductivity ($(\sigma_{xx}$, red spheres), and anomalous Hall conductivity ($ \lvert \sigma^A_{yx} \rvert$, blue spheres) as a function of temperature. (c) $ \lvert \sigma^A_{yx} \rvert$ (blue spheres) plotted as a function of $\sigma_{xx}$ in the temperature range 1.8-300 K .The solid represents a fit to Eq. \ref{scaling1}.}
    \label{hall}
    \end{center}
\end{figure*}

\begin{figure*}[!ht]
\begin{center}
\includegraphics[width=.9\linewidth]{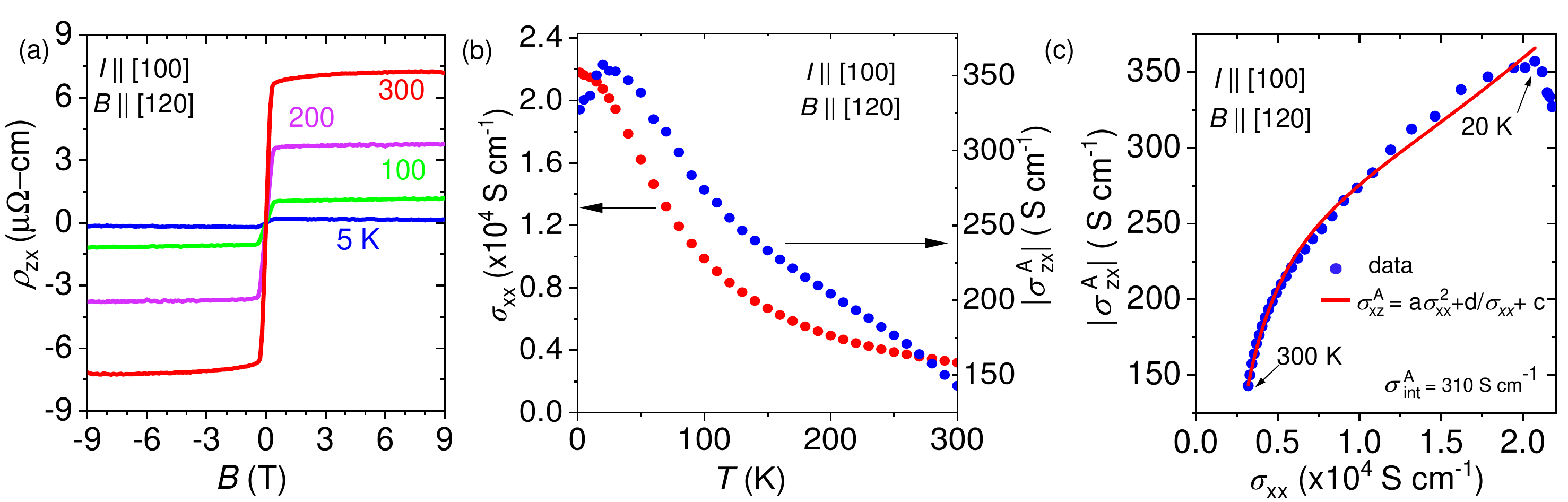}
    \caption{\small\textbf{Anomalous Hall effect of YMn$_\mathbf{6}$Sn$_\mathbf{5.45}$Ga$_\mathbf{0.55}$ measured with $\mathbf{B}~\boldsymbol{||}$ [120] and $\mathbf{I}~\boldsymbol{||}$ [100].} (a) Anomalous Hall resistivity $\rho_{zx}$ as a function of magnetic field at selected temperatures. (b) Longitudinal conductivity ($\sigma_{xx}$, red spheres), and anomalous Hall conductivity ($ \lvert \sigma^A_{xz} \lvert$, blue spheres) as a function of temperature. (c) $ \lvert \sigma^A_{zx} \rvert$ (blue spheres) plotted as a function of $\sigma_{xx}$ in the temperature range 1.8-300 K. The solid represents a fit to Eq. \ref{scaling1}.} 
    \label{hall1}
    \end{center}
\end{figure*}

Hall resistivity measured with with $B||[001]$ and the current $I||[100]$ ($\rho_{yx}$) at representative temperatures is shown in Fig. \ref{hall}(a). The zero-field value of $\rho_{yx}$ corresponds to the anomalous Hall resistivity $\rho_{yx}^{A}$. In Fig. \ref{hall}(b), we show the longitudinal Hall conductivity, $\sigma_{xx}$ = $1/\rho_{xx}$, and the absolute value of the anomalous Hall conductivity, $\sigma_{yx}^{A}$ = $-\rho_{yx}/\rho_{xx}$ (valid when $\rho_{yx}$ $\ll$ $\rho_{xx}^2$, and $\rho_{xx}$ = $\rho_{yy}$), as a function of temperature between 1.8 and 300 K. This clearly indicates that $\sigma_{yx}^{A}$ varies with temperature across the entire temperature range. The scaling of the $\sigma_{yx}^{A}$, using the relation presented in Eq. \ref{scaling1}, is shown in Fig. \ref{hall}(c). This scaling law effectively fits the high-temperature data, yielding the coefficients $a$, $d$, and $c$, as presented in Table \ref{scaling} and compared to those from Tb166. The coefficient $a$, representing impurity scattering, is about an order of magnitude larger in Y166-Ga compared to Tb166, as expected due to increased disorder scattering in the doped sample. The intrinsic anomalous Hall conductivity of Y166-Ga (121 S/cm) is comparable to that of Tb166 (140 S/cm). Interestingly, the magnitude of $d$ is smaller in the Y166-Ga than in Tb166, consistent with the hypothesis that the term $d/\sigma_{xx}$ in Eq. \ref{scaling1} is due to spin fluctuations \cite{jones2022origin}. Notably, $d\neq 0$ in Y166-Ga despite the absence of highly fluctuating Tb atoms, likely because, as discussed in  Section \ref{sec:3b}, Mn in Y166-Ga fluctuates significantly more than in Tb166. This implies that the $d/\sigma_{xx}$ is essential in the anomalous Hall scaling of $R$Mn$_6$Sn$_6$ compounds due to the presence of Mermin-Wagner fluctuations \cite{ghimire2020competing} involving either Mn or $R$ atoms.

\begin{table}[ht!]
\footnotesize
\centering
\caption{Comparison of intrinsic anomalous Hall conductivity between TbMn$_6$Sn$_6$ \cite{jones2022origin} and YMn$_6$Sn$_{5.45}$Ga$_{0.55}$.}
\resizebox{\columnwidth}{!}{%
\begin{tabular}{ |c|c|c|c| }
\hline
Compound & $c$ (S/cm) & $a$ (S/cm)$^{-1}$ & $d$ (S/cm)$^2$ \\ 
\hline
TbMn$_6$Sn$_6$ ($|\sigma_{yx}|$) & 140 & $8.69 \times 10^{-9}$ & $-2.47 \times 10^{5}$ \\ 
\hline
YMn$_6$Sn$_{5.45}$Ga$_{0.55}$ ($|\sigma_{yx}|$) & 121 & 3.25 $\times 10^{-8}$ & $-1.75 \times 10^{5} $ \\
\hline
YMn$_6$Sn$_{5.45}$Ga$_{0.55}$ ($|\sigma_{xz}|$) & 310 & $1.89\times 10^{-7}$ & $-5.41 \times 10^{5}$ \\
\hline
\end{tabular}%
}
\label{scaling}
\end{table}

The similar magnitude of intrinsic AHC in Y166-Ga and Tb166 suggests that the intrinsic AHC in $R$166 compounds is a general property of ferrimagnetism rather than a result of exotic Chern physics,  consistent with the conclusion in Ref. \cite{jones2022origin}. This interpretation is further supported by photoemission spectroscopy data from the parent compound Y166 \cite{li2021dirac}, which shows no such topological feature above the Fermi energy, with Ga doping shifting the Fermi energy even lower. 

To further validate this interpretation, we measured the Hall conductivity of Y166-Ga by applying an in-plane magnetic field, ensuring that any 2D Chern gap contributions would be excluded if present. While measuring this in-plane configuration for Tb166 would require an exceptionally high magnetic field due to magnetic saturation constraints at low temperatures \cite{kimura2006high,jones2022origin}, Y166-Ga can be measured under standard lab conditions. For consistency with the $\rho_{yx}$ measurement in Fig. \ref{hall}, we applied the current along the [100] direction and applied the in-plane field along [120]. The Hall resistivity, $\rho_{zx}$, measured across a range of temperatures from 1.8 K to 300 K, is shown in Fig. \ref{hall1}(a).

In Fig. \ref{hall1}(b), we plot $\sigma_{xx}$ = 1/$\rho_{xx}$ alongside the absolute value of the anomalous Hall conductivity, $\sigma_{zx}$ = $-\rho_{zx}/(\rho_{xx}\rho_{zz})$, which is valid when $\rho_{zx}$ $\ll$ $\rho_{xx}$ and $\rho_{zz}$. The scaling of $\sigma_{zx}$, following the relation in Eq. \ref{scaling1}, is presented in Fig. \ref{hall1}(c), with the fitting coefficients $a$, $d$, and $c$ presented in Table \ref{scaling}. Notably, the intrinsic AHC contribution, represented by coefficient $c$ ($\sigma^{A}_{zx,int}$), is significantly larger than $\sigma^{A}_{yx,int}$, indicating that the AHC in Y166-Ga has a 3D nature and can be large without invoking the Chern physics. The coefficient $|d|$ is also larger in this measurement geometry, potentially pointing to enhanced spin-fluctuations, though this cannot be directly compared to Tb166 due to differing Hall geometries and remains a topic of future work on $R$166 compounds. However, it is noteworthy that the $d/\sigma_{xx}$ term is also necessary in this case. Nevertheless, our findings reveal that the AHC in Y166-Ga displays 3D characteristics and supports theoretical calculations that suggest the AHC arises from predominantly 3D bands \cite{jones2022origin}, as further discussed in Section \ref{sec:3c}.

\subsection{Intrinsic AHC calculations for TbMn$_6$Sn$_6$ and YMn$_6$Sn$_6$ }\label{sec:3c}

\begin{figure}[!ht]
\begin{center}
\includegraphics[width=1\linewidth]{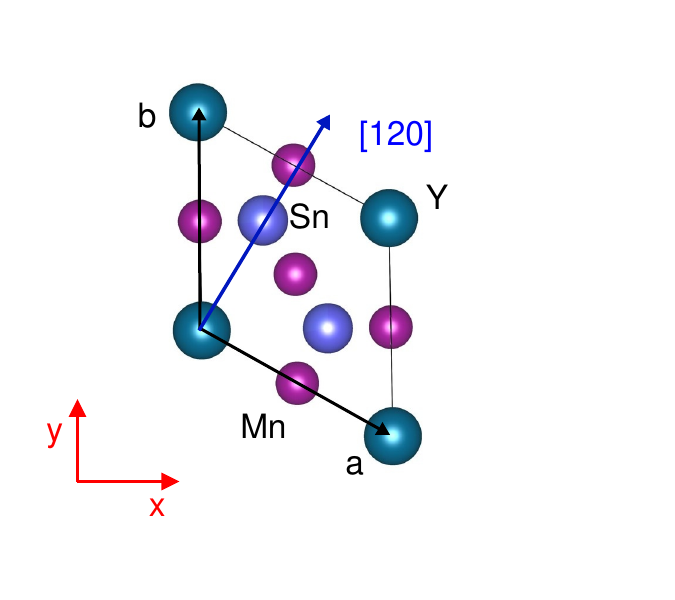}
    \caption{\small\textbf{Primitive unit cell used in AHC calculations.} The top view of the YMn$_6$Sn$_6$ primitive unit cell. The $x$- and $y$-axes are highlighted in red, while the [120] direction is denoted in blue. The $z$-axis is perpendicular to the plane.}
    \label{structure2}
    \end{center}
\end{figure}

\begin{figure}[!ht]
\centering
\begin{tabular}{c}
\includegraphics[width=.8\linewidth,clip]{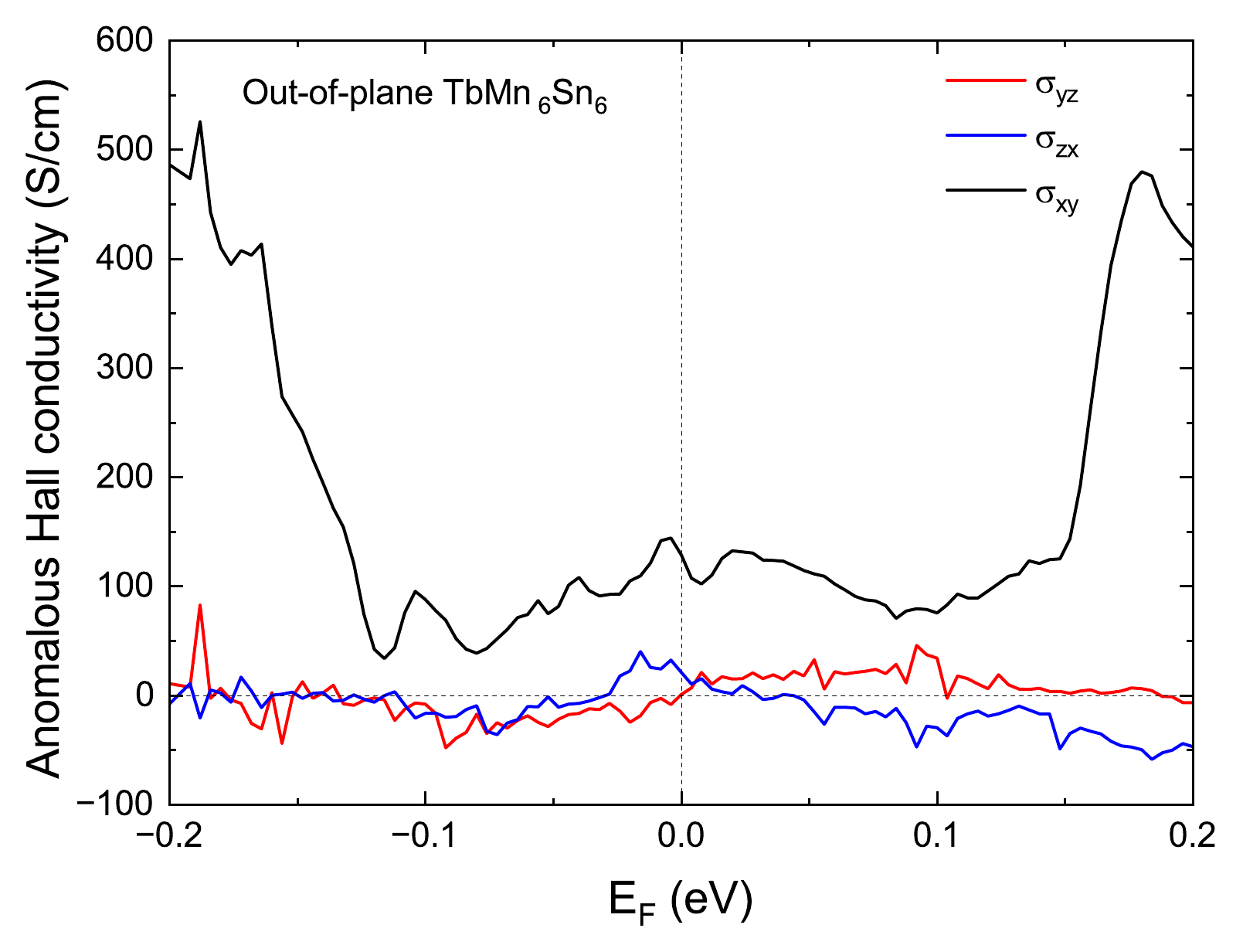} \\
\includegraphics[width=.8\linewidth,clip]{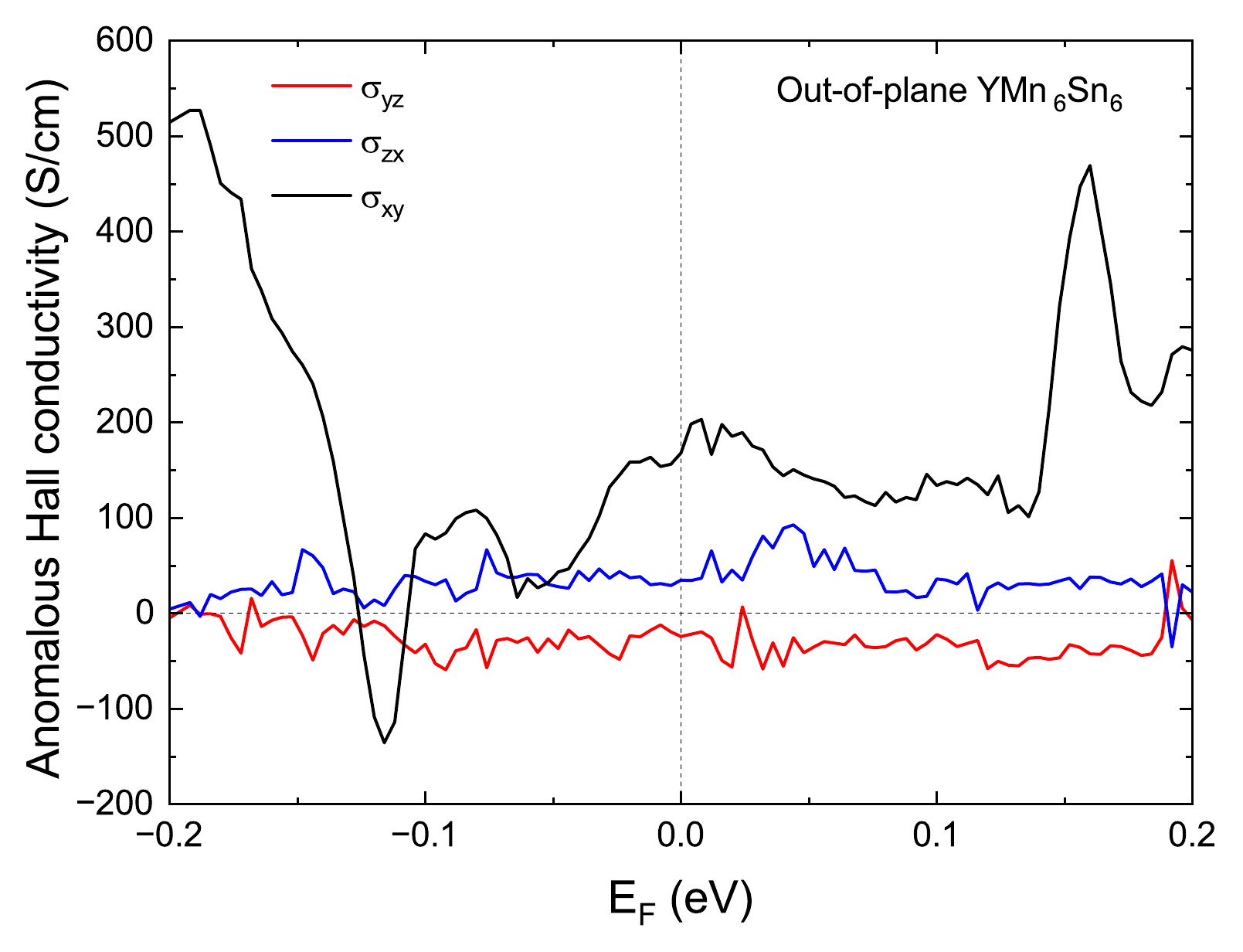} \\
\includegraphics[width=.8\linewidth,clip]{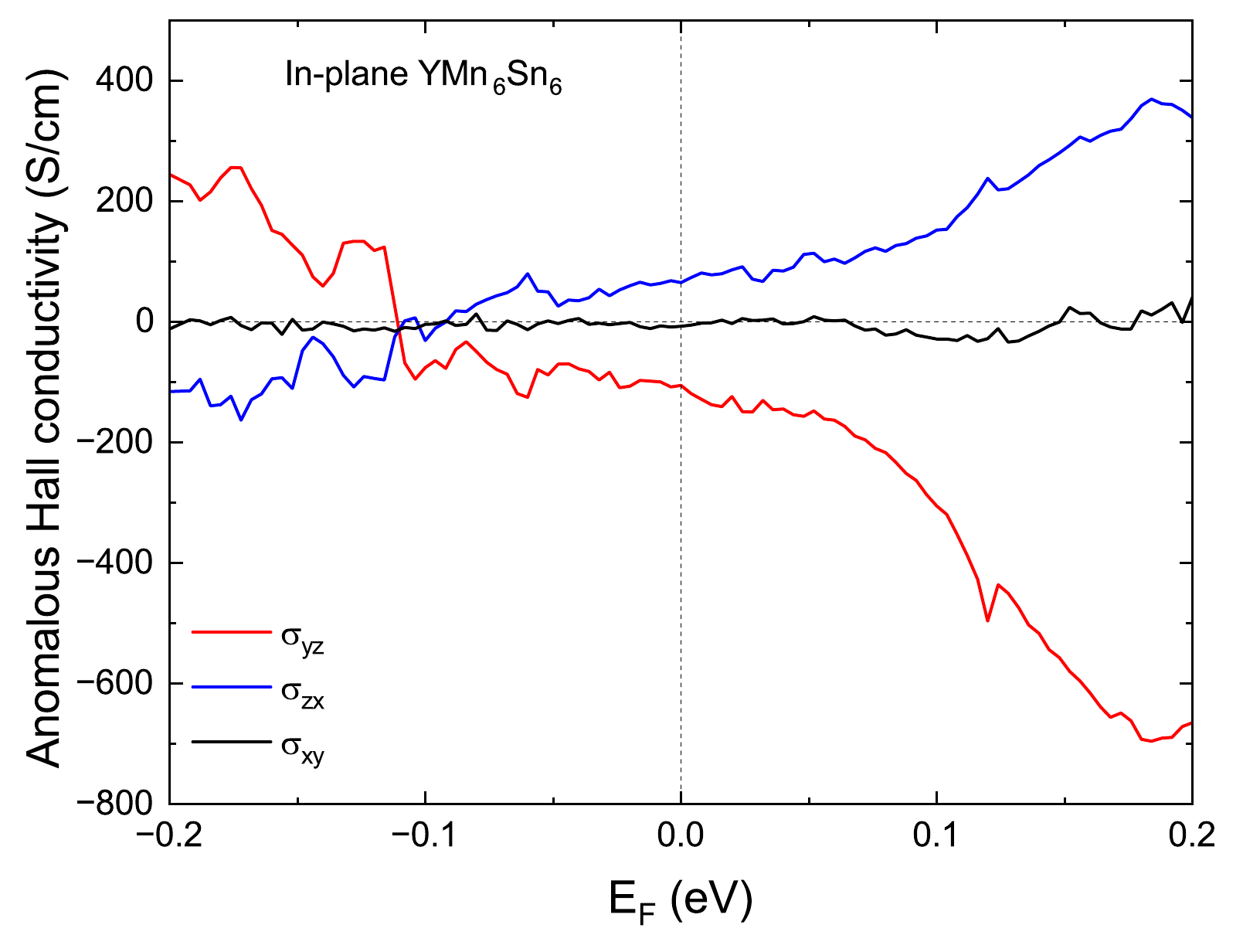} \\
\end{tabular}%
\caption{\small\textbf{Calculated intrinsic anomalous Hall conductivity (AHC) of TbMn$_6$Sn$_6$ and YMn$_6$Sn$_6$ as a function of Fermi energy ($\boldsymbol{E}_{\text{F}}$).} The intrinsic AHC of TbMn$_6$Sn$_6$ (Top) was calculated with an out-of-plane magnetization configuration, while that of YMn$_6$Sn$_6$ was calculated with both out-of-plane magnetization (Middle) and in-plane magnetization configurations (Bottom).}
\label{ymn6sn6_ahc}
\end{figure}

The intrinsic AHC can be calculated by integrating the Berry curvature over the Brillouin zone (BZ)~\cite{wang2006ab}:
\begin{eqnarray}
  \label{ahc:theo}
  \sigma_{\alpha\beta}=-\dfrac{e^2}{\hbar} \int_\text{BZ}\dfrac{d\vec{k}}{(2\pi)^3}\sum_{n} f(E_{n\vec{k}}) \Omega_{n}(\vec{k})\,
\end{eqnarray}
where $f(E_{n\vec{k}})$ is the Fermi-Dirac distribution, $\Omega_{n,\alpha\beta}(\vec{k})$ is the contribution to the Berry curvature from state $n$, and $\alpha,\beta = \{x,y,z\}$.

A notable aspect of the calculation is the pronounced and rapid oscillation observed in the Berry curvature across the BZ, requiring a dense $k$ mesh to ensure convergence.
To accelerate the calculations, we implemented Eq. \ref{ahc:theo} in our recently-developed tight-binding (TB) code~\cite{ke2019intersublattice} and carried out the AHC calculations in $R$Mn$_6$Sn${_6}$ where $R$ = Tb and Y.
A realistic TB Hamiltonian was constructed using the maximally localized Wannier functions (MLWFs) method~\cite{marzari1997maximally, souza2001maximally, marzari2012maximally} implemented in \textsc{Wannier90}~\cite{mostofi2014updated} after the self-consistent density-functional-theory calculations performed using \textsc{Wien2k}.
A set of 118 Wannier functions (WFs) consisting of Y-$4d$ (or Tb-$5d$), Mn-$3d$, and Sn-$sp$ orbitals offers an effective representation of the electronic structure near the Fermi level ($E_{\text{F}}$).
The self-consistent DFT calculations were carried out with out-of-plane magnetization in TbMn$_6$Sn${_6}$ and both in- and out-of-plane magnetization in YMn$_6$Sn${_6}$.
For the in-plane YMn$_6$Sn${_6}$ configuration, the moment is along lattice vector $\mathbf{a}$, as denoted in Fig.\ref{structure2}. 
A dense $256^3$ $k$-point mesh is used for the AHC calculations in TB.

Figure~\ref{structure2} presents the unit cell of YMn$_6$Sn$_6$ used in our calculation.
The Cartesian coordinate system is chosen so that the lattice vector $\mathbf{b}$ is along the $y$-axis, $\mathbf{c}$ is along the $z$-axis, and $\mathbf{a}$ is along the $-30^\circ$ direction off the $x$-axis.
Lattice vectors $\mathbf{a}$ and $\mathbf{b}$ point along the nearest neighboring (NN) Mn-Mn bond direction.
For the $\sigma_{xy}$ calculations discussed below, the first subscript $x$ denotes the current direction, and the second subscript $y$ denotes the Hall-field direction.

Figure~\ref{ymn6sn6_ahc} shows the AHC values calculated at $T=0$~K as functions of Fermi energy using Eq.~\ref{ahc:theo}.
In the out-of-plane orientation of both TbMn$_6$Sn$_6$ and YMn$_6$Sn$_6$, only $\sigma_{xy}$ exhibits substantial values, while $\sigma_{yz}$ and $\sigma_{zx}$ remain negligible, as illustrated in the top and middle panels of Figure~\ref{ymn6sn6_ahc}.
Conversely, with in-plane magnetization in YMn$_6$Sn$_6$, both $\sigma_{yz}$ and $\sigma_{zx}$ demonstrate appreciable values, while $\sigma_{xy}$ remains negligible, as depicted in the bottom panel of Figure~\ref{ymn6sn6_ahc}.
The band-filling calculation indicates that when doping YMn$_6$Sn$_6$ by $0.09$ hole/Mn, corresponding to a Fermi energy shift of -0.037 eV, $\sigma_{xy}=77.62$ S/cm, and $\sigma_{zx}=36.76$ S/cm.
While the former closely aligns with the experimental value, the latter is approximately 10 times smaller than the experiment. These comparison has to be taken with caution, as random substitution of 0.55 Ga likely affects the electronic structure beyond a simple rigid Fermi energy shift. Nevertheless, our theoretical calculations predict that the Hall transport in $R$166 is 3D, consistent with the experimental findings.

\section{Conclusion}\label{sec:V}

We reported results of magnetic and electrical transport measurements of YMn$_6$Sn$_{5.45}$Ga$_{0.55}$ in two different geometries, supported by first-principles and DFT calculations. Our magneto-transport measurements across these two geometries confirm a more reliable scaling law that not only extracts the intrinsic AHC, but also accounts for contributions from spin fluctuations. These measurements revealed the 3D nature of the intrinsic AHC, which we attribute to the ferromagnetic properties of the material. The excellent agreement of the AHE with the empirical scaling law over the entire temperature suggests that this scaling relation may be important for not only for the large family of $R$Mn$_6$Sn$_6$ ferro/ferrimagnetic compounds, but also for other systems with strong spin fluctuations.  

\section{ ACKNOWLEDGMENTS}
N.J.G. and H.B. acknowledge the support from the NSF CAREER award DMR-2343536. I.I.M. acknowledges support from the NSF award DMR-2403804. The work at the Ames National Laboratory was supported by the U.S. Department of Energy (US DOE), Office of Basic Energy Sciences, Division of Materials Sciences and Engineering. Ames National Laboratory is operated for the US DOE by Iowa State University under Contract No. DE-AC02-07CH11358.

Disclaimer: Certain commercial equipment, instruments, software or materials are identified in this paper in order to specify the experimental procedure adequately. Such identification is not intented to imply recommendation or endorsement by NIST, nor is it intended to imply that the materials or equipment identified are necessarily the best available for the purpose.


\section{References}


\begin{thebibliography}{54}%
\makeatletter
\providecommand \@ifxundefined [1]{%
 \@ifx{#1\undefined}
}%
\providecommand \@ifnum [1]{%
 \ifnum #1\expandafter \@firstoftwo
 \else \expandafter \@secondoftwo
 \fi
}%
\providecommand \@ifx [1]{%
 \ifx #1\expandafter \@firstoftwo
 \else \expandafter \@secondoftwo
 \fi
}%
\providecommand \natexlab [1]{#1}%
\providecommand \enquote  [1]{``#1''}%
\providecommand \bibnamefont  [1]{#1}%
\providecommand \bibfnamefont [1]{#1}%
\providecommand \citenamefont [1]{#1}%
\providecommand \href@noop [0]{\@secondoftwo}%
\providecommand \href [0]{\begingroup \@sanitize@url \@href}%
\providecommand \@href[1]{\@@startlink{#1}\@@href}%
\providecommand \@@href[1]{\endgroup#1\@@endlink}%
\providecommand \@sanitize@url [0]{\catcode `\\12\catcode `\$12\catcode `\&12\catcode `\#12\catcode `\^12\catcode `\_12\catcode `\%12\relax}%
\providecommand \@@startlink[1]{}%
\providecommand \@@endlink[0]{}%
\providecommand \url  [0]{\begingroup\@sanitize@url \@url }%
\providecommand \@url [1]{\endgroup\@href {#1}{\urlprefix }}%
\providecommand \urlprefix  [0]{URL }%
\providecommand \Eprint [0]{\href }%
\providecommand \doibase [0]{https://doi.org/}%
\providecommand \selectlanguage [0]{\@gobble}%
\providecommand \bibinfo  [0]{\@secondoftwo}%
\providecommand \bibfield  [0]{\@secondoftwo}%
\providecommand \translation [1]{[#1]}%
\providecommand \BibitemOpen [0]{}%
\providecommand \bibitemStop [0]{}%
\providecommand \bibitemNoStop [0]{.\EOS\space}%
\providecommand \EOS [0]{\spacefactor3000\relax}%
\providecommand \BibitemShut  [1]{\csname bibitem#1\endcsname}%
\let\auto@bib@innerbib\@empty
\bibitem [{\citenamefont {Zhang}(2011)}]{zhang2011quantum}%
  \BibitemOpen
  \bibfield  {author} {\bibinfo {author} {\bibfnamefont {Z.-Y.}\ \bibnamefont {Zhang}},\ }\bibfield  {title} {\bibinfo {title} {Quantum {Hall} effect in kagom{\'e} lattices under staggered magnetic field},\ }\href {https://iopscience.iop.org/article/10.1088/0953-8984/23/42/425801/meta?casa_token=M-1QRI_JpH4AAAAA:Qum_p4PVWFGv8GxRRZkTK4HBbLN3WyDQUUwfFTWotCN3gvj3wUXNof3avsIwAhQm5CMWystXg6WeCU9-E7uR7zWccF4} {\bibfield  {journal} {\bibinfo  {journal} {Journal of Physics: Condensed Matter}\ }\textbf {\bibinfo {volume} {23}},\ \bibinfo {pages} {425801} (\bibinfo {year} {2011})}\BibitemShut {NoStop}%
\bibitem [{\citenamefont {Ghimire}\ and\ \citenamefont {Mazin}(2020)}]{ghimire2020topology}%
  \BibitemOpen
  \bibfield  {author} {\bibinfo {author} {\bibfnamefont {N.~J.}\ \bibnamefont {Ghimire}}\ and\ \bibinfo {author} {\bibfnamefont {I.~I.}\ \bibnamefont {Mazin}},\ }\bibfield  {title} {\bibinfo {title} {Topology and correlations on the kagome lattice},\ }\href {https://www.nature.com/articles/s41563-019-0589-8} {\bibfield  {journal} {\bibinfo  {journal} {Nature Materials}\ }\textbf {\bibinfo {volume} {19}},\ \bibinfo {pages} {137} (\bibinfo {year} {2020})}\BibitemShut {NoStop}%
\bibitem [{\citenamefont {Kang}\ \emph {et~al.}(2020)\citenamefont {Kang}, \citenamefont {Ye}, \citenamefont {Fang}, \citenamefont {You}, \citenamefont {Levitan}, \citenamefont {Han}, \citenamefont {Facio}, \citenamefont {Jozwiak}, \citenamefont {Bostwick}, \citenamefont {Rotenberg} \emph {et~al.}}]{kang2020dirac}%
  \BibitemOpen
  \bibfield  {author} {\bibinfo {author} {\bibfnamefont {M.}~\bibnamefont {Kang}}, \bibinfo {author} {\bibfnamefont {L.}~\bibnamefont {Ye}}, \bibinfo {author} {\bibfnamefont {S.}~\bibnamefont {Fang}}, \bibinfo {author} {\bibfnamefont {J.-S.}\ \bibnamefont {You}}, \bibinfo {author} {\bibfnamefont {A.}~\bibnamefont {Levitan}}, \bibinfo {author} {\bibfnamefont {M.}~\bibnamefont {Han}}, \bibinfo {author} {\bibfnamefont {J.~I.}\ \bibnamefont {Facio}}, \bibinfo {author} {\bibfnamefont {C.}~\bibnamefont {Jozwiak}}, \bibinfo {author} {\bibfnamefont {A.}~\bibnamefont {Bostwick}}, \bibinfo {author} {\bibfnamefont {E.}~\bibnamefont {Rotenberg}}, \emph {et~al.},\ }\bibfield  {title} {\bibinfo {title} {Dirac fermions and flat bands in the ideal kagome metal {FeSn}},\ }\href {https://www.nature.com/articles/s41563-019-0531-0} {\bibfield  {journal} {\bibinfo  {journal} {Nature materials}\ }\textbf {\bibinfo {volume} {19}},\ \bibinfo {pages} {163} (\bibinfo {year} {2020})}\BibitemShut {NoStop}%
\bibitem [{\citenamefont {Ye}\ \emph {et~al.}(2018)\citenamefont {Ye}, \citenamefont {Kang}, \citenamefont {Liu}, \citenamefont {Von~Cube}, \citenamefont {Wicker}, \citenamefont {Suzuki}, \citenamefont {Jozwiak}, \citenamefont {Bostwick}, \citenamefont {Rotenberg}, \citenamefont {Bell} \emph {et~al.}}]{ye2018massive}%
  \BibitemOpen
  \bibfield  {author} {\bibinfo {author} {\bibfnamefont {L.}~\bibnamefont {Ye}}, \bibinfo {author} {\bibfnamefont {M.}~\bibnamefont {Kang}}, \bibinfo {author} {\bibfnamefont {J.}~\bibnamefont {Liu}}, \bibinfo {author} {\bibfnamefont {F.}~\bibnamefont {Von~Cube}}, \bibinfo {author} {\bibfnamefont {C.~R.}\ \bibnamefont {Wicker}}, \bibinfo {author} {\bibfnamefont {T.}~\bibnamefont {Suzuki}}, \bibinfo {author} {\bibfnamefont {C.}~\bibnamefont {Jozwiak}}, \bibinfo {author} {\bibfnamefont {A.}~\bibnamefont {Bostwick}}, \bibinfo {author} {\bibfnamefont {E.}~\bibnamefont {Rotenberg}}, \bibinfo {author} {\bibfnamefont {D.~C.}\ \bibnamefont {Bell}}, \emph {et~al.},\ }\bibfield  {title} {\bibinfo {title} {Massive {D}irac fermions in a ferromagnetic kagome metal},\ }\href {https://www.nature.com/articles/nature25987} {\bibfield  {journal} {\bibinfo  {journal} {Nature}\ }\textbf {\bibinfo {volume} {555}},\ \bibinfo {pages} {638} (\bibinfo {year} {2018})}\BibitemShut {NoStop}%
\bibitem [{\citenamefont {Yin}\ \emph {et~al.}(2019)\citenamefont {Yin}, \citenamefont {Zhang}, \citenamefont {Chang}, \citenamefont {Wang}, \citenamefont {Tsirkin}, \citenamefont {Guguchia}, \citenamefont {Lian}, \citenamefont {Zhou}, \citenamefont {Jiang}, \citenamefont {Belopolski} \emph {et~al.}}]{yin2019negative}%
  \BibitemOpen
  \bibfield  {author} {\bibinfo {author} {\bibfnamefont {J.-X.}\ \bibnamefont {Yin}}, \bibinfo {author} {\bibfnamefont {S.~S.}\ \bibnamefont {Zhang}}, \bibinfo {author} {\bibfnamefont {G.}~\bibnamefont {Chang}}, \bibinfo {author} {\bibfnamefont {Q.}~\bibnamefont {Wang}}, \bibinfo {author} {\bibfnamefont {S.~S.}\ \bibnamefont {Tsirkin}}, \bibinfo {author} {\bibfnamefont {Z.}~\bibnamefont {Guguchia}}, \bibinfo {author} {\bibfnamefont {B.}~\bibnamefont {Lian}}, \bibinfo {author} {\bibfnamefont {H.}~\bibnamefont {Zhou}}, \bibinfo {author} {\bibfnamefont {K.}~\bibnamefont {Jiang}}, \bibinfo {author} {\bibfnamefont {I.}~\bibnamefont {Belopolski}}, \emph {et~al.},\ }\bibfield  {title} {\bibinfo {title} {Negative flat band magnetism in a spin--orbit-coupled correlated kagome magnet},\ }\href {https://www.nature.com/articles/s41567-019-0426-7} {\bibfield  {journal} {\bibinfo  {journal} {Nature Physics}\ }\textbf {\bibinfo {volume} {15}},\ \bibinfo {pages} {443} (\bibinfo {year} {2019})}\BibitemShut {NoStop}%
\bibitem [{\citenamefont {Kuroda}\ \emph {et~al.}(2017)\citenamefont {Kuroda}, \citenamefont {Tomita}, \citenamefont {Suzuki}, \citenamefont {Bareille}, \citenamefont {Nugroho}, \citenamefont {Goswami}, \citenamefont {Ochi}, \citenamefont {Ikhlas}, \citenamefont {Nakayama}, \citenamefont {Akebi} \emph {et~al.}}]{kuroda2017evidence}%
  \BibitemOpen
  \bibfield  {author} {\bibinfo {author} {\bibfnamefont {K.}~\bibnamefont {Kuroda}}, \bibinfo {author} {\bibfnamefont {T.}~\bibnamefont {Tomita}}, \bibinfo {author} {\bibfnamefont {M.-T.}\ \bibnamefont {Suzuki}}, \bibinfo {author} {\bibfnamefont {C.}~\bibnamefont {Bareille}}, \bibinfo {author} {\bibfnamefont {A.}~\bibnamefont {Nugroho}}, \bibinfo {author} {\bibfnamefont {P.}~\bibnamefont {Goswami}}, \bibinfo {author} {\bibfnamefont {M.}~\bibnamefont {Ochi}}, \bibinfo {author} {\bibfnamefont {M.}~\bibnamefont {Ikhlas}}, \bibinfo {author} {\bibfnamefont {M.}~\bibnamefont {Nakayama}}, \bibinfo {author} {\bibfnamefont {S.}~\bibnamefont {Akebi}}, \emph {et~al.},\ }\bibfield  {title} {\bibinfo {title} {Evidence for magnetic {W}eyl fermions in a correlated metal},\ }\href {https://www.nature.com/articles/nmat4987} {\bibfield  {journal} {\bibinfo  {journal} {Nature materials}\ }\textbf {\bibinfo {volume} {16}},\ \bibinfo {pages} {1090} (\bibinfo {year} {2017})}\BibitemShut {NoStop}%
\bibitem [{\citenamefont {Mazin}\ \emph {et~al.}(2014)\citenamefont {Mazin}, \citenamefont {Jeschke}, \citenamefont {Lechermann}, \citenamefont {Lee}, \citenamefont {Fink}, \citenamefont {Thomale},\ and\ \citenamefont {Valent{\'i}}}]{superconductivity}%
  \BibitemOpen
  \bibfield  {author} {\bibinfo {author} {\bibfnamefont {I.~I.}\ \bibnamefont {Mazin}}, \bibinfo {author} {\bibfnamefont {H.~O.}\ \bibnamefont {Jeschke}}, \bibinfo {author} {\bibfnamefont {F.}~\bibnamefont {Lechermann}}, \bibinfo {author} {\bibfnamefont {H.}~\bibnamefont {Lee}}, \bibinfo {author} {\bibfnamefont {M.}~\bibnamefont {Fink}}, \bibinfo {author} {\bibfnamefont {R.}~\bibnamefont {Thomale}},\ and\ \bibinfo {author} {\bibfnamefont {R.}~\bibnamefont {Valent{\'i}}},\ }\bibfield  {title} {\bibinfo {title} {Theoretical prediction of a strongly correlated {D}irac metal},\ }\href {https://doi.org/10.1038/ncomms5261} {\bibfield  {journal} {\bibinfo  {journal} {Nature Communications}\ }\textbf {\bibinfo {volume} {5}},\ \bibinfo {pages} {4261} (\bibinfo {year} {2014})}\BibitemShut {NoStop}%
\bibitem [{\citenamefont {Bolens}\ and\ \citenamefont {Nagaosa}(2019)}]{bolens2019topological}%
  \BibitemOpen
  \bibfield  {author} {\bibinfo {author} {\bibfnamefont {A.}~\bibnamefont {Bolens}}\ and\ \bibinfo {author} {\bibfnamefont {N.}~\bibnamefont {Nagaosa}},\ }\bibfield  {title} {\bibinfo {title} {Topological states on the breathing kagome lattice},\ }\href {https://journals.aps.org/prb/abstract/10.1103/PhysRevB.99.165141} {\bibfield  {journal} {\bibinfo  {journal} {Physical Review B}\ }\textbf {\bibinfo {volume} {99}},\ \bibinfo {pages} {165141} (\bibinfo {year} {2019})}\BibitemShut {NoStop}%
\bibitem [{\citenamefont {Asaba}\ \emph {et~al.}(2020)\citenamefont {Asaba}, \citenamefont {Thomas}, \citenamefont {Curtis}, \citenamefont {Thompson}, \citenamefont {Bauer},\ and\ \citenamefont {Ronning}}]{asaba2020anomalous}%
  \BibitemOpen
  \bibfield  {author} {\bibinfo {author} {\bibfnamefont {T.}~\bibnamefont {Asaba}}, \bibinfo {author} {\bibfnamefont {S.~M.}\ \bibnamefont {Thomas}}, \bibinfo {author} {\bibfnamefont {M.}~\bibnamefont {Curtis}}, \bibinfo {author} {\bibfnamefont {J.~D.}\ \bibnamefont {Thompson}}, \bibinfo {author} {\bibfnamefont {E.~D.}\ \bibnamefont {Bauer}},\ and\ \bibinfo {author} {\bibfnamefont {F.}~\bibnamefont {Ronning}},\ }\bibfield  {title} {\bibinfo {title} {Anomalous {H}all effect in the kagome ferrimagnet {GdMn$_6$Sn$_6$}},\ }\href {https://journals.aps.org/prb/abstract/10.1103/PhysRevB.101.174415} {\bibfield  {journal} {\bibinfo  {journal} {Physical Review B}\ }\textbf {\bibinfo {volume} {101}},\ \bibinfo {pages} {174415} (\bibinfo {year} {2020})}\BibitemShut {NoStop}%
\bibitem [{\citenamefont {Bhandari}\ \emph {et~al.}(2024)\citenamefont {Bhandari}, \citenamefont {Dally}, \citenamefont {Siegfried}, \citenamefont {Regmi}, \citenamefont {Rule}, \citenamefont {Chi}, \citenamefont {Lynn}, \citenamefont {Mazin},\ and\ \citenamefont {Ghimire}}]{bhandari2024magnetism}%
  \BibitemOpen
  \bibfield  {author} {\bibinfo {author} {\bibfnamefont {H.}~\bibnamefont {Bhandari}}, \bibinfo {author} {\bibfnamefont {R.~L.}\ \bibnamefont {Dally}}, \bibinfo {author} {\bibfnamefont {P.~E.}\ \bibnamefont {Siegfried}}, \bibinfo {author} {\bibfnamefont {R.~B.}\ \bibnamefont {Regmi}}, \bibinfo {author} {\bibfnamefont {K.~C.}\ \bibnamefont {Rule}}, \bibinfo {author} {\bibfnamefont {S.}~\bibnamefont {Chi}}, \bibinfo {author} {\bibfnamefont {J.~W.}\ \bibnamefont {Lynn}}, \bibinfo {author} {\bibfnamefont {I.}~\bibnamefont {Mazin}},\ and\ \bibinfo {author} {\bibfnamefont {N.~J.}\ \bibnamefont {Ghimire}},\ }\bibfield  {title} {\bibinfo {title} {Magnetism and fermiology of kagome magnet {YMn$_6$Sn$_4$Ge$_2$}},\ }\href {https://www.nature.com/articles/s41535-023-00616-0} {\bibfield  {journal} {\bibinfo  {journal} {npj Quantum Materials}\ }\textbf {\bibinfo {volume} {9}},\ \bibinfo {pages} {6} (\bibinfo {year} {2024})}\BibitemShut {NoStop}%
\bibitem [{\citenamefont {Pokharel}\ \emph {et~al.}(2021)\citenamefont {Pokharel}, \citenamefont {Teicher}, \citenamefont {Ortiz}, \citenamefont {Sarte}, \citenamefont {Wu}, \citenamefont {Peng}, \citenamefont {He}, \citenamefont {Seshadri},\ and\ \citenamefont {Wilson}}]{pokharel2021electronic}%
  \BibitemOpen
  \bibfield  {author} {\bibinfo {author} {\bibfnamefont {G.}~\bibnamefont {Pokharel}}, \bibinfo {author} {\bibfnamefont {S.~M.}\ \bibnamefont {Teicher}}, \bibinfo {author} {\bibfnamefont {B.~R.}\ \bibnamefont {Ortiz}}, \bibinfo {author} {\bibfnamefont {P.~M.}\ \bibnamefont {Sarte}}, \bibinfo {author} {\bibfnamefont {G.}~\bibnamefont {Wu}}, \bibinfo {author} {\bibfnamefont {S.}~\bibnamefont {Peng}}, \bibinfo {author} {\bibfnamefont {J.}~\bibnamefont {He}}, \bibinfo {author} {\bibfnamefont {R.}~\bibnamefont {Seshadri}},\ and\ \bibinfo {author} {\bibfnamefont {S.~D.}\ \bibnamefont {Wilson}},\ }\bibfield  {title} {\bibinfo {title} {Electronic properties of the topological kagome metals {YV$_6$Sn$_6$} and {GdV$_6$Sn$_6$}},\ }\href {https://journals.aps.org/prb/abstract/10.1103/PhysRevB.104.235139} {\bibfield  {journal} {\bibinfo  {journal} {Physical Review B}\ }\textbf {\bibinfo {volume} {104}},\ \bibinfo {pages} {235139} (\bibinfo {year} {2021})}\BibitemShut {NoStop}%
\bibitem [{\citenamefont {Jones}\ \emph {et~al.}(2024)\citenamefont {Jones}, \citenamefont {Das}, \citenamefont {Bhandari}, \citenamefont {Liu}, \citenamefont {Siegfried}, \citenamefont {Ghimire}, \citenamefont {Tsirkin}, \citenamefont {Mazin},\ and\ \citenamefont {Ghimire}}]{jones2022origin}%
  \BibitemOpen
  \bibfield  {author} {\bibinfo {author} {\bibfnamefont {D.~C.}\ \bibnamefont {Jones}}, \bibinfo {author} {\bibfnamefont {S.}~\bibnamefont {Das}}, \bibinfo {author} {\bibfnamefont {H.}~\bibnamefont {Bhandari}}, \bibinfo {author} {\bibfnamefont {X.}~\bibnamefont {Liu}}, \bibinfo {author} {\bibfnamefont {P.}~\bibnamefont {Siegfried}}, \bibinfo {author} {\bibfnamefont {M.~P.}\ \bibnamefont {Ghimire}}, \bibinfo {author} {\bibfnamefont {S.~S.}\ \bibnamefont {Tsirkin}}, \bibinfo {author} {\bibfnamefont {I.~I.}\ \bibnamefont {Mazin}},\ and\ \bibinfo {author} {\bibfnamefont {N.~J.}\ \bibnamefont {Ghimire}},\ }\bibfield  {title} {\bibinfo {title} {Origin of spin reorientation and intrinsic anomalous {H}all effect in the kagome ferrimagnet {TbMn$_6$Sn$_6$}},\ }\href {https://doi.org/10.1103/PhysRevB.110.115134} {\bibfield  {journal} {\bibinfo  {journal} {Phys. Rev. B}\ }\textbf {\bibinfo {volume} {110}},\ \bibinfo {pages} {115134} (\bibinfo {year} {2024})}\BibitemShut {NoStop}%
\bibitem [{\citenamefont {Pokharel}\ \emph {et~al.}(2022)\citenamefont {Pokharel}, \citenamefont {Ortiz}, \citenamefont {Chamorro}, \citenamefont {Sarte}, \citenamefont {Kautzsch}, \citenamefont {Wu}, \citenamefont {Ruff},\ and\ \citenamefont {Wilson}}]{pokharel2022highly}%
  \BibitemOpen
  \bibfield  {author} {\bibinfo {author} {\bibfnamefont {G.}~\bibnamefont {Pokharel}}, \bibinfo {author} {\bibfnamefont {B.}~\bibnamefont {Ortiz}}, \bibinfo {author} {\bibfnamefont {J.}~\bibnamefont {Chamorro}}, \bibinfo {author} {\bibfnamefont {P.}~\bibnamefont {Sarte}}, \bibinfo {author} {\bibfnamefont {L.}~\bibnamefont {Kautzsch}}, \bibinfo {author} {\bibfnamefont {G.}~\bibnamefont {Wu}}, \bibinfo {author} {\bibfnamefont {J.}~\bibnamefont {Ruff}},\ and\ \bibinfo {author} {\bibfnamefont {S.~D.}\ \bibnamefont {Wilson}},\ }\bibfield  {title} {\bibinfo {title} {Highly anisotropic magnetism in the vanadium-based kagome metal {TbV$_6$Sn$_6$}},\ }\href {https://journals.aps.org/prmaterials/abstract/10.1103/PhysRevMaterials.6.104202} {\bibfield  {journal} {\bibinfo  {journal} {Physical Review Materials}\ }\textbf {\bibinfo {volume} {6}},\ \bibinfo {pages} {104202} (\bibinfo {year} {2022})}\BibitemShut {NoStop}%
\bibitem [{\citenamefont {Arachchige}\ \emph {et~al.}(2022)\citenamefont {Arachchige}, \citenamefont {Meier}, \citenamefont {Marshall}, \citenamefont {Matsuoka}, \citenamefont {Xue}, \citenamefont {McGuire}, \citenamefont {Hermann}, \citenamefont {Cao},\ and\ \citenamefont {Mandrus}}]{arachchige2022charge}%
  \BibitemOpen
  \bibfield  {author} {\bibinfo {author} {\bibfnamefont {H.~W.~S.}\ \bibnamefont {Arachchige}}, \bibinfo {author} {\bibfnamefont {W.~R.}\ \bibnamefont {Meier}}, \bibinfo {author} {\bibfnamefont {M.}~\bibnamefont {Marshall}}, \bibinfo {author} {\bibfnamefont {T.}~\bibnamefont {Matsuoka}}, \bibinfo {author} {\bibfnamefont {R.}~\bibnamefont {Xue}}, \bibinfo {author} {\bibfnamefont {M.~A.}\ \bibnamefont {McGuire}}, \bibinfo {author} {\bibfnamefont {R.~P.}\ \bibnamefont {Hermann}}, \bibinfo {author} {\bibfnamefont {H.}~\bibnamefont {Cao}},\ and\ \bibinfo {author} {\bibfnamefont {D.}~\bibnamefont {Mandrus}},\ }\bibfield  {title} {\bibinfo {title} {Charge {D}ensity {W}ave in {K}agome {L}attice {I}ntermetallic {ScV$_6$Sn$_6$}},\ }\href {https://journals.aps.org/prl/abstract/10.1103/PhysRevLett.129.216402} {\bibfield  {journal} {\bibinfo  {journal} {Physical Review Letters}\ }\textbf {\bibinfo {volume} {129}},\ \bibinfo {pages} {216402} (\bibinfo {year} {2022})}\BibitemShut {NoStop}%
\bibitem [{\citenamefont {Riberolles}\ \emph {et~al.}(2023)\citenamefont {Riberolles}, \citenamefont {Slade}, \citenamefont {Dally}, \citenamefont {Sarte}, \citenamefont {Li}, \citenamefont {Han}, \citenamefont {Lane}, \citenamefont {Stock}, \citenamefont {Bhandari}, \citenamefont {Ghimire} \emph {et~al.}}]{riberolles2023orbital}%
  \BibitemOpen
  \bibfield  {author} {\bibinfo {author} {\bibfnamefont {S.}~\bibnamefont {Riberolles}}, \bibinfo {author} {\bibfnamefont {T.~J.}\ \bibnamefont {Slade}}, \bibinfo {author} {\bibfnamefont {R.}~\bibnamefont {Dally}}, \bibinfo {author} {\bibfnamefont {P.}~\bibnamefont {Sarte}}, \bibinfo {author} {\bibfnamefont {B.}~\bibnamefont {Li}}, \bibinfo {author} {\bibfnamefont {T.}~\bibnamefont {Han}}, \bibinfo {author} {\bibfnamefont {H.}~\bibnamefont {Lane}}, \bibinfo {author} {\bibfnamefont {C.}~\bibnamefont {Stock}}, \bibinfo {author} {\bibfnamefont {H.}~\bibnamefont {Bhandari}}, \bibinfo {author} {\bibfnamefont {N.}~\bibnamefont {Ghimire}}, \emph {et~al.},\ }\bibfield  {title} {\bibinfo {title} {Orbital character of the spin-reorientation transition in {TbMn$_6$Sn$_6$}},\ }\href {https://www.nature.com/articles/s41467-023-38174-5} {\bibfield  {journal} {\bibinfo  {journal} {Nature Communications}\ }\textbf {\bibinfo {volume} {14}},\ \bibinfo {pages} {2658} (\bibinfo {year} {2023})}\BibitemShut {NoStop}%
\bibitem [{\citenamefont {Lee}\ \emph {et~al.}(2023)\citenamefont {Lee}, \citenamefont {Skomski}, \citenamefont {Wang}, \citenamefont {Orth}, \citenamefont {Ren}, \citenamefont {Kang}, \citenamefont {Pathak}, \citenamefont {Kutepov}, \citenamefont {Harmon}, \citenamefont {McQueeney}, \citenamefont {Mazin},\ and\ \citenamefont {Ke}}]{lee2022interplay}%
  \BibitemOpen
  \bibfield  {author} {\bibinfo {author} {\bibfnamefont {Y.}~\bibnamefont {Lee}}, \bibinfo {author} {\bibfnamefont {R.}~\bibnamefont {Skomski}}, \bibinfo {author} {\bibfnamefont {X.}~\bibnamefont {Wang}}, \bibinfo {author} {\bibfnamefont {P.~P.}\ \bibnamefont {Orth}}, \bibinfo {author} {\bibfnamefont {Y.}~\bibnamefont {Ren}}, \bibinfo {author} {\bibfnamefont {B.}~\bibnamefont {Kang}}, \bibinfo {author} {\bibfnamefont {A.~K.}\ \bibnamefont {Pathak}}, \bibinfo {author} {\bibfnamefont {A.}~\bibnamefont {Kutepov}}, \bibinfo {author} {\bibfnamefont {B.~N.}\ \bibnamefont {Harmon}}, \bibinfo {author} {\bibfnamefont {R.~J.}\ \bibnamefont {McQueeney}}, \bibinfo {author} {\bibfnamefont {I.~I.}\ \bibnamefont {Mazin}},\ and\ \bibinfo {author} {\bibfnamefont {L.}~\bibnamefont {Ke}},\ }\bibfield  {title} {\bibinfo {title} {Interplay between magnetism and band topology in the kagome magnets {RMn$_6$Sn$_6$}},\ }\href {https://doi.org/10.1103/PhysRevB.108.045132} {\bibfield  {journal} {\bibinfo  {journal} {Phys. Rev. B}\
  }\textbf {\bibinfo {volume} {108}},\ \bibinfo {pages} {045132} (\bibinfo {year} {2023})}\BibitemShut {NoStop}%
\bibitem [{\citenamefont {Ghimire}\ \emph {et~al.}(2020)\citenamefont {Ghimire}, \citenamefont {Dally}, \citenamefont {Poudel}, \citenamefont {Jones}, \citenamefont {Michel}, \citenamefont {Magar}, \citenamefont {Bleuel}, \citenamefont {McGuire}, \citenamefont {Jiang}, \citenamefont {Mitchell} \emph {et~al.}}]{ghimire2020competing}%
  \BibitemOpen
  \bibfield  {author} {\bibinfo {author} {\bibfnamefont {N.~J.}\ \bibnamefont {Ghimire}}, \bibinfo {author} {\bibfnamefont {R.~L.}\ \bibnamefont {Dally}}, \bibinfo {author} {\bibfnamefont {L.}~\bibnamefont {Poudel}}, \bibinfo {author} {\bibfnamefont {D.}~\bibnamefont {Jones}}, \bibinfo {author} {\bibfnamefont {D.}~\bibnamefont {Michel}}, \bibinfo {author} {\bibfnamefont {N.~T.}\ \bibnamefont {Magar}}, \bibinfo {author} {\bibfnamefont {M.}~\bibnamefont {Bleuel}}, \bibinfo {author} {\bibfnamefont {M.~A.}\ \bibnamefont {McGuire}}, \bibinfo {author} {\bibfnamefont {J.}~\bibnamefont {Jiang}}, \bibinfo {author} {\bibfnamefont {J.}~\bibnamefont {Mitchell}}, \emph {et~al.},\ }\bibfield  {title} {\bibinfo {title} {Competing magnetic phases and fluctuation-driven scalar spin chirality in the kagome metal {YMn$_6$Sn$_6$}},\ }\href {https://www.science.org/doi/full/10.1126/sciadv.abe2680} {\bibfield  {journal} {\bibinfo  {journal} {Science Advances}\ }\textbf {\bibinfo {volume} {6}},\ \bibinfo {pages} {eabe2680}
  (\bibinfo {year} {2020})}\BibitemShut {NoStop}%
\bibitem [{\citenamefont {Wang}\ \emph {et~al.}(2021)\citenamefont {Wang}, \citenamefont {Neubauer}, \citenamefont {Duan}, \citenamefont {Yin}, \citenamefont {Fujitsu}, \citenamefont {Hosono}, \citenamefont {Ye}, \citenamefont {Zhang}, \citenamefont {Chi}, \citenamefont {Krycka} \emph {et~al.}}]{wang2021field}%
  \BibitemOpen
  \bibfield  {author} {\bibinfo {author} {\bibfnamefont {Q.}~\bibnamefont {Wang}}, \bibinfo {author} {\bibfnamefont {K.~J.}\ \bibnamefont {Neubauer}}, \bibinfo {author} {\bibfnamefont {C.}~\bibnamefont {Duan}}, \bibinfo {author} {\bibfnamefont {Q.}~\bibnamefont {Yin}}, \bibinfo {author} {\bibfnamefont {S.}~\bibnamefont {Fujitsu}}, \bibinfo {author} {\bibfnamefont {H.}~\bibnamefont {Hosono}}, \bibinfo {author} {\bibfnamefont {F.}~\bibnamefont {Ye}}, \bibinfo {author} {\bibfnamefont {R.}~\bibnamefont {Zhang}}, \bibinfo {author} {\bibfnamefont {S.}~\bibnamefont {Chi}}, \bibinfo {author} {\bibfnamefont {K.}~\bibnamefont {Krycka}}, \emph {et~al.},\ }\bibfield  {title} {\bibinfo {title} {Field-induced topological {H}all effect and double-fan spin structure with a c-axis component in the metallic kagome antiferromagnetic compound {YMn$_6$Sn$_6$}},\ }\href {https://journals.aps.org/prb/abstract/10.1103/PhysRevB.103.014416} {\bibfield  {journal} {\bibinfo  {journal} {Physical Review B}\ }\textbf {\bibinfo {volume}
  {103}},\ \bibinfo {pages} {014416} (\bibinfo {year} {2021})}\BibitemShut {NoStop}%
\bibitem [{\citenamefont {Siegfried}\ \emph {et~al.}(2022)\citenamefont {Siegfried}, \citenamefont {Bhandari}, \citenamefont {Jones}, \citenamefont {Ghimire}, \citenamefont {Dally}, \citenamefont {Poudel}, \citenamefont {Bleuel}, \citenamefont {Lynn}, \citenamefont {Mazin},\ and\ \citenamefont {Ghimire}}]{siegfried2022magnetization}%
  \BibitemOpen
  \bibfield  {author} {\bibinfo {author} {\bibfnamefont {P.~E.}\ \bibnamefont {Siegfried}}, \bibinfo {author} {\bibfnamefont {H.}~\bibnamefont {Bhandari}}, \bibinfo {author} {\bibfnamefont {D.~C.}\ \bibnamefont {Jones}}, \bibinfo {author} {\bibfnamefont {M.~P.}\ \bibnamefont {Ghimire}}, \bibinfo {author} {\bibfnamefont {R.~L.}\ \bibnamefont {Dally}}, \bibinfo {author} {\bibfnamefont {L.}~\bibnamefont {Poudel}}, \bibinfo {author} {\bibfnamefont {M.}~\bibnamefont {Bleuel}}, \bibinfo {author} {\bibfnamefont {J.~W.}\ \bibnamefont {Lynn}}, \bibinfo {author} {\bibfnamefont {I.~I.}\ \bibnamefont {Mazin}},\ and\ \bibinfo {author} {\bibfnamefont {N.~J.}\ \bibnamefont {Ghimire}},\ }\bibfield  {title} {\bibinfo {title} {Magnetization-driven {L}ifshitz transition and charge-spin coupling in the kagome metal {YMn$_6$Sn$_6$}},\ }\href {https://www.nature.com/articles/s42005-022-00833-2} {\bibfield  {journal} {\bibinfo  {journal} {Communications Physics}\ }\textbf {\bibinfo {volume} {5}},\ \bibinfo {pages} {1} (\bibinfo
  {year} {2022})}\BibitemShut {NoStop}%
\bibitem [{\citenamefont {Yin}\ \emph {et~al.}(2020)\citenamefont {Yin}, \citenamefont {Ma}, \citenamefont {Cochran}, \citenamefont {Xu}, \citenamefont {Zhang}, \citenamefont {Tien}, \citenamefont {Shumiya}, \citenamefont {Cheng}, \citenamefont {Jiang}, \citenamefont {Lian} \emph {et~al.}}]{yin2020quantum}%
  \BibitemOpen
  \bibfield  {author} {\bibinfo {author} {\bibfnamefont {J.-X.}\ \bibnamefont {Yin}}, \bibinfo {author} {\bibfnamefont {W.}~\bibnamefont {Ma}}, \bibinfo {author} {\bibfnamefont {T.~A.}\ \bibnamefont {Cochran}}, \bibinfo {author} {\bibfnamefont {X.}~\bibnamefont {Xu}}, \bibinfo {author} {\bibfnamefont {S.~S.}\ \bibnamefont {Zhang}}, \bibinfo {author} {\bibfnamefont {H.-J.}\ \bibnamefont {Tien}}, \bibinfo {author} {\bibfnamefont {N.}~\bibnamefont {Shumiya}}, \bibinfo {author} {\bibfnamefont {G.}~\bibnamefont {Cheng}}, \bibinfo {author} {\bibfnamefont {K.}~\bibnamefont {Jiang}}, \bibinfo {author} {\bibfnamefont {B.}~\bibnamefont {Lian}}, \emph {et~al.},\ }\bibfield  {title} {\bibinfo {title} {Quantum-limit {C}hern topological magnetism in {TbMn$_6$Sn$_6$}},\ }\href {https://www.nature.com/articles/s41586-020-2482-7} {\bibfield  {journal} {\bibinfo  {journal} {Nature}\ }\textbf {\bibinfo {volume} {583}},\ \bibinfo {pages} {533} (\bibinfo {year} {2020})}\BibitemShut {NoStop}%
\bibitem [{\citenamefont {Riberolles}\ \emph {et~al.}(2022)\citenamefont {Riberolles}, \citenamefont {Slade}, \citenamefont {Abernathy}, \citenamefont {Granroth}, \citenamefont {Li}, \citenamefont {Lee}, \citenamefont {Canfield}, \citenamefont {Ueland}, \citenamefont {Ke},\ and\ \citenamefont {McQueeney}}]{riberolles2022low}%
  \BibitemOpen
  \bibfield  {author} {\bibinfo {author} {\bibfnamefont {S.}~\bibnamefont {Riberolles}}, \bibinfo {author} {\bibfnamefont {T.~J.}\ \bibnamefont {Slade}}, \bibinfo {author} {\bibfnamefont {D.}~\bibnamefont {Abernathy}}, \bibinfo {author} {\bibfnamefont {G.}~\bibnamefont {Granroth}}, \bibinfo {author} {\bibfnamefont {B.}~\bibnamefont {Li}}, \bibinfo {author} {\bibfnamefont {Y.}~\bibnamefont {Lee}}, \bibinfo {author} {\bibfnamefont {P.}~\bibnamefont {Canfield}}, \bibinfo {author} {\bibfnamefont {B.~G.}\ \bibnamefont {Ueland}}, \bibinfo {author} {\bibfnamefont {L.}~\bibnamefont {Ke}},\ and\ \bibinfo {author} {\bibfnamefont {R.~J.}\ \bibnamefont {McQueeney}},\ }\bibfield  {title} {\bibinfo {title} {Low-{T}emperature {C}ompeting {M}agnetic {E}nergy {S}cales in the {T}opological {F}errimagnet {TbMn$_6$Sn$_6$}},\ }\href {https://journals.aps.org/prx/abstract/10.1103/PhysRevX.12.021043} {\bibfield  {journal} {\bibinfo  {journal} {Physical Review X}\ }\textbf {\bibinfo {volume} {12}},\ \bibinfo {pages} {021043}
  (\bibinfo {year} {2022})}\BibitemShut {NoStop}%
\bibitem [{\citenamefont {Mielke~III}\ \emph {et~al.}(2022)\citenamefont {Mielke~III}, \citenamefont {Ma}, \citenamefont {Pomjakushin}, \citenamefont {Zaharko}, \citenamefont {Sturniolo}, \citenamefont {Liu}, \citenamefont {Ukleev}, \citenamefont {White}, \citenamefont {Yin}, \citenamefont {Tsirkin} \emph {et~al.}}]{mielke2022low}%
  \BibitemOpen
  \bibfield  {author} {\bibinfo {author} {\bibfnamefont {C.}~\bibnamefont {Mielke~III}}, \bibinfo {author} {\bibfnamefont {W.}~\bibnamefont {Ma}}, \bibinfo {author} {\bibfnamefont {V.}~\bibnamefont {Pomjakushin}}, \bibinfo {author} {\bibfnamefont {O.}~\bibnamefont {Zaharko}}, \bibinfo {author} {\bibfnamefont {S.}~\bibnamefont {Sturniolo}}, \bibinfo {author} {\bibfnamefont {X.}~\bibnamefont {Liu}}, \bibinfo {author} {\bibfnamefont {V.}~\bibnamefont {Ukleev}}, \bibinfo {author} {\bibfnamefont {J.}~\bibnamefont {White}}, \bibinfo {author} {\bibfnamefont {J.-X.}\ \bibnamefont {Yin}}, \bibinfo {author} {\bibfnamefont {S.}~\bibnamefont {Tsirkin}}, \emph {et~al.},\ }\bibfield  {title} {\bibinfo {title} {Low-temperature magnetic crossover in the topological kagome magnet {TbMn$_6$Sn$_6$}},\ }\href {https://www.nature.com/articles/s42005-022-00885-4} {\bibfield  {journal} {\bibinfo  {journal} {Communications Physics}\ }\textbf {\bibinfo {volume} {5}},\ \bibinfo {pages} {107} (\bibinfo {year} {2022})}\BibitemShut
  {NoStop}%
\bibitem [{\citenamefont {Li}\ \emph {et~al.}(2023)\citenamefont {Li}, \citenamefont {Yin}, \citenamefont {Jiang}, \citenamefont {Zhu}, \citenamefont {Gao}, \citenamefont {Wang}, \citenamefont {Shen}, \citenamefont {Zhao}, \citenamefont {Cai}, \citenamefont {Lei} \emph {et~al.}}]{li2023discovery}%
  \BibitemOpen
  \bibfield  {author} {\bibinfo {author} {\bibfnamefont {Z.}~\bibnamefont {Li}}, \bibinfo {author} {\bibfnamefont {Q.}~\bibnamefont {Yin}}, \bibinfo {author} {\bibfnamefont {Y.}~\bibnamefont {Jiang}}, \bibinfo {author} {\bibfnamefont {Z.}~\bibnamefont {Zhu}}, \bibinfo {author} {\bibfnamefont {Y.}~\bibnamefont {Gao}}, \bibinfo {author} {\bibfnamefont {S.}~\bibnamefont {Wang}}, \bibinfo {author} {\bibfnamefont {J.}~\bibnamefont {Shen}}, \bibinfo {author} {\bibfnamefont {T.}~\bibnamefont {Zhao}}, \bibinfo {author} {\bibfnamefont {J.}~\bibnamefont {Cai}}, \bibinfo {author} {\bibfnamefont {H.}~\bibnamefont {Lei}}, \emph {et~al.},\ }\bibfield  {title} {\bibinfo {title} {Discovery of {T}opological {M}agnetic {T}extures near {R}oom {T}emperature in {Q}uantum {M}agnet {TbMn$_6$Sn$_6$}},\ }\href {https://onlinelibrary.wiley.com/doi/full/10.1002/adma.202211164?casa_token=K49xq7voNeIAAAAA%3AD7nD8zN0xKrjhm0xnQOKHj-T5L12LjwhpF6rJCvwJVhzqn79Hi55TD1VyVcUaHBI1GwTO0gEliKS-A} {\bibfield  {journal} {\bibinfo  {journal} {Advanced
  Materials}\ ,\ \bibinfo {pages} {2211164}} (\bibinfo {year} {2023})}\BibitemShut {NoStop}%
\bibitem [{\citenamefont {Xu}\ \emph {et~al.}(2022)\citenamefont {Xu}, \citenamefont {Yin}, \citenamefont {Ma}, \citenamefont {Tien}, \citenamefont {Qiang}, \citenamefont {Reddy}, \citenamefont {Zhou}, \citenamefont {Shen}, \citenamefont {Lu}, \citenamefont {Chang} \emph {et~al.}}]{xu2022topological}%
  \BibitemOpen
  \bibfield  {author} {\bibinfo {author} {\bibfnamefont {X.}~\bibnamefont {Xu}}, \bibinfo {author} {\bibfnamefont {J.-X.}\ \bibnamefont {Yin}}, \bibinfo {author} {\bibfnamefont {W.}~\bibnamefont {Ma}}, \bibinfo {author} {\bibfnamefont {H.-J.}\ \bibnamefont {Tien}}, \bibinfo {author} {\bibfnamefont {X.-B.}\ \bibnamefont {Qiang}}, \bibinfo {author} {\bibfnamefont {P.~S.}\ \bibnamefont {Reddy}}, \bibinfo {author} {\bibfnamefont {H.}~\bibnamefont {Zhou}}, \bibinfo {author} {\bibfnamefont {J.}~\bibnamefont {Shen}}, \bibinfo {author} {\bibfnamefont {H.-Z.}\ \bibnamefont {Lu}}, \bibinfo {author} {\bibfnamefont {T.-R.}\ \bibnamefont {Chang}}, \emph {et~al.},\ }\bibfield  {title} {\bibinfo {title} {Topological charge-entropy scaling in kagome {C}hern magnet {TbMn$_6$Sn$_6$}},\ }\href {https://www.nature.com/articles/s41467-022-28796-6} {\bibfield  {journal} {\bibinfo  {journal} {Nature communications}\ }\textbf {\bibinfo {volume} {13}},\ \bibinfo {pages} {1197} (\bibinfo {year} {2022})}\BibitemShut {NoStop}%
\bibitem [{\citenamefont {Gao}\ \emph {et~al.}(2021)\citenamefont {Gao}, \citenamefont {Shen}, \citenamefont {Wang}, \citenamefont {Shi}, \citenamefont {Zhao}, \citenamefont {Li}, \citenamefont {Cao}, \citenamefont {Pei}, \citenamefont {Ge}, \citenamefont {Li} \emph {et~al.}}]{gao2021anomalous}%
  \BibitemOpen
  \bibfield  {author} {\bibinfo {author} {\bibfnamefont {L.}~\bibnamefont {Gao}}, \bibinfo {author} {\bibfnamefont {S.}~\bibnamefont {Shen}}, \bibinfo {author} {\bibfnamefont {Q.}~\bibnamefont {Wang}}, \bibinfo {author} {\bibfnamefont {W.}~\bibnamefont {Shi}}, \bibinfo {author} {\bibfnamefont {Y.}~\bibnamefont {Zhao}}, \bibinfo {author} {\bibfnamefont {C.}~\bibnamefont {Li}}, \bibinfo {author} {\bibfnamefont {W.}~\bibnamefont {Cao}}, \bibinfo {author} {\bibfnamefont {C.}~\bibnamefont {Pei}}, \bibinfo {author} {\bibfnamefont {J.-Y.}\ \bibnamefont {Ge}}, \bibinfo {author} {\bibfnamefont {G.}~\bibnamefont {Li}}, \emph {et~al.},\ }\bibfield  {title} {\bibinfo {title} {Anomalous {H}all effect in ferrimagnetic metal {RMn$_6$Sn$_6$ (R= Tb, Dy, Ho) with clean Mn kagome lattice}},\ }\href {https://pubs.aip.org/aip/apl/article-abstract/119/9/092405/41513/Anomalous-Hall-effect-in-ferrimagnetic-metal?redirectedFrom=fulltext} {\bibfield  {journal} {\bibinfo  {journal} {Applied Physics Letters}\ }\textbf {\bibinfo {volume}
  {119}} (\bibinfo {year} {2021})}\BibitemShut {NoStop}%
\bibitem [{\citenamefont {Wenzel}\ \emph {et~al.}(2022)\citenamefont {Wenzel}, \citenamefont {Tsirlin}, \citenamefont {Iakutkina}, \citenamefont {Yin}, \citenamefont {Lei}, \citenamefont {Dressel},\ and\ \citenamefont {Uykur}}]{Wenzel2022Effect}%
  \BibitemOpen
  \bibfield  {author} {\bibinfo {author} {\bibfnamefont {M.}~\bibnamefont {Wenzel}}, \bibinfo {author} {\bibfnamefont {A.~A.}\ \bibnamefont {Tsirlin}}, \bibinfo {author} {\bibfnamefont {O.}~\bibnamefont {Iakutkina}}, \bibinfo {author} {\bibfnamefont {Q.}~\bibnamefont {Yin}}, \bibinfo {author} {\bibfnamefont {H.~C.}\ \bibnamefont {Lei}}, \bibinfo {author} {\bibfnamefont {M.}~\bibnamefont {Dressel}},\ and\ \bibinfo {author} {\bibfnamefont {E.}~\bibnamefont {Uykur}},\ }\bibfield  {title} {\bibinfo {title} {Effect of magnetism and phonons on localized carriers in the ferrimagnetic kagome metals \text{GdMn}$_{6}$\text{Sn}$_{6}$ and \text{TbMn}$_{6}$\text{Sn}$_{6}$},\ }\href {https://doi.org/10.1103/PhysRevB.106.L241108} {\bibfield  {journal} {\bibinfo  {journal} {Phys. Rev. B}\ }\textbf {\bibinfo {volume} {106}},\ \bibinfo {pages} {L241108} (\bibinfo {year} {2022})}\BibitemShut {NoStop}%
\bibitem [{\citenamefont {Zhang}\ \emph {et~al.}(2022{\natexlab{a}})\citenamefont {Zhang}, \citenamefont {Koo}, \citenamefont {Xu}, \citenamefont {Sretenovic}, \citenamefont {Yan},\ and\ \citenamefont {Ke}}]{zhang2022exchange}%
  \BibitemOpen
  \bibfield  {author} {\bibinfo {author} {\bibfnamefont {H.}~\bibnamefont {Zhang}}, \bibinfo {author} {\bibfnamefont {J.}~\bibnamefont {Koo}}, \bibinfo {author} {\bibfnamefont {C.}~\bibnamefont {Xu}}, \bibinfo {author} {\bibfnamefont {M.}~\bibnamefont {Sretenovic}}, \bibinfo {author} {\bibfnamefont {B.}~\bibnamefont {Yan}},\ and\ \bibinfo {author} {\bibfnamefont {X.}~\bibnamefont {Ke}},\ }\bibfield  {title} {\bibinfo {title} {Exchange-biased topological transverse thermoelectric effects in a {K}agome ferrimagnet},\ }\href {https://www.nature.com/articles/s41467-022-28733-7} {\bibfield  {journal} {\bibinfo  {journal} {Nature communications}\ }\textbf {\bibinfo {volume} {13}},\ \bibinfo {pages} {1091} (\bibinfo {year} {2022}{\natexlab{a}})}\BibitemShut {NoStop}%
\bibitem [{\citenamefont {Cr\'epieux}\ and\ \citenamefont {Bruno}(2001)}]{PhysRevB.64.014416}%
  \BibitemOpen
  \bibfield  {author} {\bibinfo {author} {\bibfnamefont {A.}~\bibnamefont {Cr\'epieux}}\ and\ \bibinfo {author} {\bibfnamefont {P.}~\bibnamefont {Bruno}},\ }\bibfield  {title} {\bibinfo {title} {Theory of the anomalous {H}all effect from the {K}ubo formula and the {D}irac equation},\ }\href {https://doi.org/10.1103/PhysRevB.64.014416} {\bibfield  {journal} {\bibinfo  {journal} {Phys. Rev. B}\ }\textbf {\bibinfo {volume} {64}},\ \bibinfo {pages} {014416} (\bibinfo {year} {2001})}\BibitemShut {NoStop}%
\bibitem [{\citenamefont {Tian}\ \emph {et~al.}(2009)\citenamefont {Tian}, \citenamefont {Ye},\ and\ \citenamefont {Jin}}]{Tian2009}%
  \BibitemOpen
  \bibfield  {author} {\bibinfo {author} {\bibfnamefont {Y.}~\bibnamefont {Tian}}, \bibinfo {author} {\bibfnamefont {L.}~\bibnamefont {Ye}},\ and\ \bibinfo {author} {\bibfnamefont {X.}~\bibnamefont {Jin}},\ }\bibfield  {title} {\bibinfo {title} {Proper {S}caling of the {A}nomalous {H}all {E}ffect},\ }\href {https://doi.org/10.1103/PhysRevLett.103.087206} {\bibfield  {journal} {\bibinfo  {journal} {Phys. Rev. Lett.}\ }\textbf {\bibinfo {volume} {103}},\ \bibinfo {pages} {087206} (\bibinfo {year} {2009})}\BibitemShut {NoStop}%
\bibitem [{\citenamefont {Grigoryan}\ \emph {et~al.}(2017)\citenamefont {Grigoryan}, \citenamefont {Xiao}, \citenamefont {Wang},\ and\ \citenamefont {Xia}}]{PhysRevB.96.144426}%
  \BibitemOpen
  \bibfield  {author} {\bibinfo {author} {\bibfnamefont {V.~L.}\ \bibnamefont {Grigoryan}}, \bibinfo {author} {\bibfnamefont {J.}~\bibnamefont {Xiao}}, \bibinfo {author} {\bibfnamefont {X.}~\bibnamefont {Wang}},\ and\ \bibinfo {author} {\bibfnamefont {K.}~\bibnamefont {Xia}},\ }\bibfield  {title} {\bibinfo {title} {Anomalous {H}all effect scaling in ferromagnetic thin films},\ }\href {https://doi.org/10.1103/PhysRevB.96.144426} {\bibfield  {journal} {\bibinfo  {journal} {Phys. Rev. B}\ }\textbf {\bibinfo {volume} {96}},\ \bibinfo {pages} {144426} (\bibinfo {year} {2017})}\BibitemShut {NoStop}%
\bibitem [{\citenamefont {Xu}\ \emph {et~al.}(2021)\citenamefont {Xu}, \citenamefont {Heitmann}, \citenamefont {Zhang}, \citenamefont {Xu},\ and\ \citenamefont {Ke}}]{xu2021magnetic}%
  \BibitemOpen
  \bibfield  {author} {\bibinfo {author} {\bibfnamefont {C.}~\bibnamefont {Xu}}, \bibinfo {author} {\bibfnamefont {T.}~\bibnamefont {Heitmann}}, \bibinfo {author} {\bibfnamefont {H.}~\bibnamefont {Zhang}}, \bibinfo {author} {\bibfnamefont {X.}~\bibnamefont {Xu}},\ and\ \bibinfo {author} {\bibfnamefont {X.}~\bibnamefont {Ke}},\ }\bibfield  {title} {\bibinfo {title} {Magnetic phase transition, magnetoresistance, and anomalous {H}all effect in {Ga-substituted YMn$_6$Sn$_6$ with a ferromagnetic kagome lattice}},\ }\href {https://journals.aps.org/prb/abstract/10.1103/PhysRevB.104.024413} {\bibfield  {journal} {\bibinfo  {journal} {Physical Review B}\ }\textbf {\bibinfo {volume} {104}},\ \bibinfo {pages} {024413} (\bibinfo {year} {2021})}\BibitemShut {NoStop}%
\bibitem [{\citenamefont {{Bruker AXS}}(2021)}]{apex4}%
  \BibitemOpen
  \bibfield  {author} {\bibinfo {author} {\bibnamefont {{Bruker AXS}}},\ }\href@noop {} {\bibinfo {title} {{APEX-4}}},\ \bibinfo {howpublished} {Bruker AXS, Madison, Wisconsin, USA} (\bibinfo {year} {2021})\BibitemShut {NoStop}%
\bibitem [{\citenamefont {Krause}\ \emph {et~al.}(2015)\citenamefont {Krause}, \citenamefont {Herbst-Irmer}, \citenamefont {Sheldrick},\ and\ \citenamefont {Stalke}}]{krause2015comparison}%
  \BibitemOpen
  \bibfield  {author} {\bibinfo {author} {\bibfnamefont {L.}~\bibnamefont {Krause}}, \bibinfo {author} {\bibfnamefont {R.}~\bibnamefont {Herbst-Irmer}}, \bibinfo {author} {\bibfnamefont {G.~M.}\ \bibnamefont {Sheldrick}},\ and\ \bibinfo {author} {\bibfnamefont {D.}~\bibnamefont {Stalke}},\ }\bibfield  {title} {\bibinfo {title} {Comparison of silver and molybdenum microfocus {X}-ray sources for single-crystal structure determination},\ }\href {https://journals.iucr.org/j/issues/2015/01/00/aj5242/index.html} {\bibfield  {journal} {\bibinfo  {journal} {Journal of applied crystallography}\ }\textbf {\bibinfo {volume} {48}},\ \bibinfo {pages} {3} (\bibinfo {year} {2015})}\BibitemShut {NoStop}%
\bibitem [{\citenamefont {Sheldrick}(2015{\natexlab{a}})}]{sheldrick2015shelxt}%
  \BibitemOpen
  \bibfield  {author} {\bibinfo {author} {\bibfnamefont {G.~M.}\ \bibnamefont {Sheldrick}},\ }\bibfield  {title} {\bibinfo {title} {Shelxt--integrated space-group and crystal-structure determination},\ }\href {https://journals.iucr.org/a/issues/2015/01/00/sc5086/sc5086.pdf} {\bibfield  {journal} {\bibinfo  {journal} {Acta Crystallographica Section A: Foundations and Advances}\ }\textbf {\bibinfo {volume} {71}},\ \bibinfo {pages} {3} (\bibinfo {year} {2015}{\natexlab{a}})}\BibitemShut {NoStop}%
\bibitem [{\citenamefont {Sheldrick}(2015{\natexlab{b}})}]{sheldrick2015crystal}%
  \BibitemOpen
  \bibfield  {author} {\bibinfo {author} {\bibfnamefont {G.~M.}\ \bibnamefont {Sheldrick}},\ }\bibfield  {title} {\bibinfo {title} {Crystal structure refinement with shelxl},\ }\href {https://journals.iucr.org/c/issues/2015/01/00/fa3356/fa3356.pdf} {\bibfield  {journal} {\bibinfo  {journal} {Acta Crystallographica Section C: Structural Chemistry}\ }\textbf {\bibinfo {volume} {71}},\ \bibinfo {pages} {3} (\bibinfo {year} {2015}{\natexlab{b}})}\BibitemShut {NoStop}%
\bibitem [{\citenamefont {Kresse}\ and\ \citenamefont {Furthm\"uller}(1996)}]{Kresse1996}%
  \BibitemOpen
  \bibfield  {author} {\bibinfo {author} {\bibfnamefont {G.}~\bibnamefont {Kresse}}\ and\ \bibinfo {author} {\bibfnamefont {J.}~\bibnamefont {Furthm\"uller}},\ }\bibfield  {title} {\bibinfo {title} {Efficient iterative schemes for ab initio total-energy calculations using a plane-wave basis set},\ }\href {https://doi.org/10.1103/PhysRevB.54.11169} {\bibfield  {journal} {\bibinfo  {journal} {Phys. Rev. B}\ }\textbf {\bibinfo {volume} {54}},\ \bibinfo {pages} {11169} (\bibinfo {year} {1996})}\BibitemShut {NoStop}%
\bibitem [{\citenamefont {Bl\"ochl}(1994)}]{blochl1994}%
  \BibitemOpen
  \bibfield  {author} {\bibinfo {author} {\bibfnamefont {P.~E.}\ \bibnamefont {Bl\"ochl}},\ }\bibfield  {title} {\bibinfo {title} {Projector augmented-wave method},\ }\href {https://doi.org/10.1103/PhysRevB.50.17953} {\bibfield  {journal} {\bibinfo  {journal} {Phys. Rev. B}\ }\textbf {\bibinfo {volume} {50}},\ \bibinfo {pages} {17953} (\bibinfo {year} {1994})}\BibitemShut {NoStop}%
\bibitem [{\citenamefont {Kresse}\ and\ \citenamefont {Joubert}(1999)}]{Kresse1999}%
  \BibitemOpen
  \bibfield  {author} {\bibinfo {author} {\bibfnamefont {G.}~\bibnamefont {Kresse}}\ and\ \bibinfo {author} {\bibfnamefont {D.}~\bibnamefont {Joubert}},\ }\bibfield  {title} {\bibinfo {title} {From ultrasoft pseudopotentials to the projector augmented-wave method},\ }\href {https://doi.org/10.1103/PhysRevB.59.1758} {\bibfield  {journal} {\bibinfo  {journal} {Phys. Rev. B}\ }\textbf {\bibinfo {volume} {59}},\ \bibinfo {pages} {1758} (\bibinfo {year} {1999})}\BibitemShut {NoStop}%
\bibitem [{\citenamefont {Perdew}\ \emph {et~al.}(1996)\citenamefont {Perdew}, \citenamefont {Burke},\ and\ \citenamefont {Ernzerhof}}]{PBE}%
  \BibitemOpen
  \bibfield  {author} {\bibinfo {author} {\bibfnamefont {J.~P.}\ \bibnamefont {Perdew}}, \bibinfo {author} {\bibfnamefont {K.}~\bibnamefont {Burke}},\ and\ \bibinfo {author} {\bibfnamefont {M.}~\bibnamefont {Ernzerhof}},\ }\bibfield  {title} {\bibinfo {title} {Generalized {G}radient {A}pproximation {M}ade {S}imple},\ }\href {https://doi.org/10.1103/PhysRevLett.77.3865} {\bibfield  {journal} {\bibinfo  {journal} {Phys. Rev. Lett.}\ }\textbf {\bibinfo {volume} {77}},\ \bibinfo {pages} {3865} (\bibinfo {year} {1996})}\BibitemShut {NoStop}%
\bibitem [{\citenamefont {Anisimov}\ and\ \citenamefont {Gunnarsson}(1991)}]{Anisimov1991}%
  \BibitemOpen
  \bibfield  {author} {\bibinfo {author} {\bibfnamefont {V.~I.}\ \bibnamefont {Anisimov}}\ and\ \bibinfo {author} {\bibfnamefont {O.}~\bibnamefont {Gunnarsson}},\ }\bibfield  {title} {\bibinfo {title} {Density-functional calculation of effective {C}oulomb interactions in metals},\ }\href {https://doi.org/10.1103/PhysRevB.43.7570} {\bibfield  {journal} {\bibinfo  {journal} {Phys. Rev. B}\ }\textbf {\bibinfo {volume} {43}},\ \bibinfo {pages} {7570} (\bibinfo {year} {1991})}\BibitemShut {NoStop}%
\bibitem [{\citenamefont {Zhang}\ \emph {et~al.}(2001)\citenamefont {Zhang}, \citenamefont {Zhao}, \citenamefont {Cheng}, \citenamefont {Li}, \citenamefont {Sun}, \citenamefont {Zhang},\ and\ \citenamefont {Shen}}]{zhang2001magnetism}%
  \BibitemOpen
  \bibfield  {author} {\bibinfo {author} {\bibfnamefont {S.-y.}\ \bibnamefont {Zhang}}, \bibinfo {author} {\bibfnamefont {P.}~\bibnamefont {Zhao}}, \bibinfo {author} {\bibfnamefont {Z.-h.}\ \bibnamefont {Cheng}}, \bibinfo {author} {\bibfnamefont {R.-w.}\ \bibnamefont {Li}}, \bibinfo {author} {\bibfnamefont {J.-r.}\ \bibnamefont {Sun}}, \bibinfo {author} {\bibfnamefont {H.-w.}\ \bibnamefont {Zhang}},\ and\ \bibinfo {author} {\bibfnamefont {B.-g.}\ \bibnamefont {Shen}},\ }\bibfield  {title} {\bibinfo {title} {Magnetism and giant magnetoresistance of {YMn$_6$Sn$_{6-x}$Ga$_x$ (x= 0-1.8) compounds}},\ }\href {https://journals.aps.org/prb/abstract/10.1103/PhysRevB.64.212404} {\bibfield  {journal} {\bibinfo  {journal} {Physical Review B}\ }\textbf {\bibinfo {volume} {64}},\ \bibinfo {pages} {212404} (\bibinfo {year} {2001})}\BibitemShut {NoStop}%
\bibitem [{\citenamefont {Dally}\ \emph {et~al.}(2021)\citenamefont {Dally}, \citenamefont {Lynn}, \citenamefont {Ghimire}, \citenamefont {Michel}, \citenamefont {Siegfried},\ and\ \citenamefont {Mazin}}]{dally2021chiral}%
  \BibitemOpen
  \bibfield  {author} {\bibinfo {author} {\bibfnamefont {R.~L.}\ \bibnamefont {Dally}}, \bibinfo {author} {\bibfnamefont {J.~W.}\ \bibnamefont {Lynn}}, \bibinfo {author} {\bibfnamefont {N.~J.}\ \bibnamefont {Ghimire}}, \bibinfo {author} {\bibfnamefont {D.}~\bibnamefont {Michel}}, \bibinfo {author} {\bibfnamefont {P.}~\bibnamefont {Siegfried}},\ and\ \bibinfo {author} {\bibfnamefont {I.~I.}\ \bibnamefont {Mazin}},\ }\bibfield  {title} {\bibinfo {title} {Chiral properties of the zero-field spiral state and field-induced magnetic phases of the itinerant kagome metal {YMn$_6$Sn$_6$}},\ }\href {https://journals.aps.org/prb/abstract/10.1103/PhysRevB.103.094413} {\bibfield  {journal} {\bibinfo  {journal} {Physical Review B}\ }\textbf {\bibinfo {volume} {103}},\ \bibinfo {pages} {094413} (\bibinfo {year} {2021})}\BibitemShut {NoStop}%
\bibitem [{\citenamefont {Idrissi}\ \emph {et~al.}(1991)\citenamefont {Idrissi}, \citenamefont {Venturini},\ and\ \citenamefont {Malaman}}]{Idrissi1991}%
  \BibitemOpen
  \bibfield  {author} {\bibinfo {author} {\bibfnamefont {B.~C.~E.}\ \bibnamefont {Idrissi}}, \bibinfo {author} {\bibfnamefont {G.}~\bibnamefont {Venturini}},\ and\ \bibinfo {author} {\bibfnamefont {B.}~\bibnamefont {Malaman}},\ }\bibfield  {title} {\bibinfo {title} {{Magnetic structures of TbMn$_6$Sn$_6$ and HoMn$_6$Sn$_6$ compounds from neutron diffraction study}},\ }\href {https://doi.org/10.1016/0022-5088(91)90359-C} {\bibfield  {journal} {\bibinfo  {journal} {Journal of the Less Common Metals}\ }\textbf {\bibinfo {volume} {175}},\ \bibinfo {pages} {143} (\bibinfo {year} {1991})}\BibitemShut {NoStop}%
\bibitem [{\citenamefont {Rosenfeld}\ and\ \citenamefont {Mushnikov}(2008)}]{ROSENFELD20081898}%
  \BibitemOpen
  \bibfield  {author} {\bibinfo {author} {\bibfnamefont {E.}~\bibnamefont {Rosenfeld}}\ and\ \bibinfo {author} {\bibfnamefont {N.}~\bibnamefont {Mushnikov}},\ }\bibfield  {title} {\bibinfo {title} {Double-flat-spiral magnetic structures: Theory and application to the {RMn$_6$X$_6$} compounds},\ }\href {https://doi.org/https://doi.org/10.1016/j.physb.2007.10.220} {\bibfield  {journal} {\bibinfo  {journal} {Physica B: Condensed Matter}\ }\textbf {\bibinfo {volume} {403}},\ \bibinfo {pages} {1898} (\bibinfo {year} {2008})}\BibitemShut {NoStop}%
\bibitem [{\citenamefont {{P. Blaha}}\ \emph {et~al.}(2002)\citenamefont {{P. Blaha}}, \citenamefont {{K. Schwarz}}, \citenamefont {{G. K. H. Madsen}}, \citenamefont {{D. Kvasnicka}},\ and\ \citenamefont {{J. Luitz}}}]{WIEN}%
  \BibitemOpen
  \bibfield  {author} {\bibinfo {author} {\bibnamefont {{P. Blaha}}}, \bibinfo {author} {\bibnamefont {{K. Schwarz}}}, \bibinfo {author} {\bibnamefont {{G. K. H. Madsen}}}, \bibinfo {author} {\bibnamefont {{D. Kvasnicka}}},\ and\ \bibinfo {author} {\bibnamefont {{J. Luitz}}},\ }\href@noop {} {\bibinfo {title} {{WIEN2K}}} (\bibinfo {year} {2002}),\ \bibinfo {note} {{ISBN 3-9501031-1-2}}\BibitemShut {NoStop}%
\bibitem [{\citenamefont {Zhang}\ \emph {et~al.}(2022{\natexlab{b}})\citenamefont {Zhang}, \citenamefont {Koo}, \citenamefont {Xu}, \citenamefont {Sretenovic}, \citenamefont {Yan},\ and\ \citenamefont {Ke}}]{Binghai}%
  \BibitemOpen
  \bibfield  {author} {\bibinfo {author} {\bibfnamefont {H.}~\bibnamefont {Zhang}}, \bibinfo {author} {\bibfnamefont {J.}~\bibnamefont {Koo}}, \bibinfo {author} {\bibfnamefont {C.}~\bibnamefont {Xu}}, \bibinfo {author} {\bibfnamefont {M.}~\bibnamefont {Sretenovic}}, \bibinfo {author} {\bibfnamefont {B.}~\bibnamefont {Yan}},\ and\ \bibinfo {author} {\bibfnamefont {X.}~\bibnamefont {Ke}},\ }\bibfield  {title} {\bibinfo {title} {{Exchange-biased topological transverse thermoelectric effects in a Kagome ferrimagnet}},\ }\href {https://doi.org/10.1038/s41467-022-28733-7} {\bibfield  {journal} {\bibinfo  {journal} {Nature Communications}\ }\textbf {\bibinfo {volume} {13}},\ \bibinfo {pages} {1091} (\bibinfo {year} {2022}{\natexlab{b}})}\BibitemShut {NoStop}%
\bibitem [{\citenamefont {Li}\ \emph {et~al.}(2021)\citenamefont {Li}, \citenamefont {Wang}, \citenamefont {Wang}, \citenamefont {Yuan}, \citenamefont {Song}, \citenamefont {Lou}, \citenamefont {Liu}, \citenamefont {Huang}, \citenamefont {Liu}, \citenamefont {Lei} \emph {et~al.}}]{li2021dirac}%
  \BibitemOpen
  \bibfield  {author} {\bibinfo {author} {\bibfnamefont {M.}~\bibnamefont {Li}}, \bibinfo {author} {\bibfnamefont {Q.}~\bibnamefont {Wang}}, \bibinfo {author} {\bibfnamefont {G.}~\bibnamefont {Wang}}, \bibinfo {author} {\bibfnamefont {Z.}~\bibnamefont {Yuan}}, \bibinfo {author} {\bibfnamefont {W.}~\bibnamefont {Song}}, \bibinfo {author} {\bibfnamefont {R.}~\bibnamefont {Lou}}, \bibinfo {author} {\bibfnamefont {Z.}~\bibnamefont {Liu}}, \bibinfo {author} {\bibfnamefont {Y.}~\bibnamefont {Huang}}, \bibinfo {author} {\bibfnamefont {Z.}~\bibnamefont {Liu}}, \bibinfo {author} {\bibfnamefont {H.}~\bibnamefont {Lei}}, \emph {et~al.},\ }\bibfield  {title} {\bibinfo {title} {Dirac cone, flat band and saddle point in kagome magnet {YMn$_6$Sn$_6$}},\ }\href {https://www.nature.com/articles/s41467-021-23536-8} {\bibfield  {journal} {\bibinfo  {journal} {Nature communications}\ }\textbf {\bibinfo {volume} {12}},\ \bibinfo {pages} {1} (\bibinfo {year} {2021})}\BibitemShut {NoStop}%
\bibitem [{\citenamefont {Kimura}\ \emph {et~al.}(2006)\citenamefont {Kimura}, \citenamefont {Matsuo}, \citenamefont {Yoshii}, \citenamefont {Kindo}, \citenamefont {Zhang}, \citenamefont {Br{\"u}ck}, \citenamefont {Buschow}, \citenamefont {De~Boer}, \citenamefont {Lef{\`e}vre},\ and\ \citenamefont {Venturini}}]{kimura2006high}%
  \BibitemOpen
  \bibfield  {author} {\bibinfo {author} {\bibfnamefont {S.}~\bibnamefont {Kimura}}, \bibinfo {author} {\bibfnamefont {A.}~\bibnamefont {Matsuo}}, \bibinfo {author} {\bibfnamefont {S.}~\bibnamefont {Yoshii}}, \bibinfo {author} {\bibfnamefont {K.}~\bibnamefont {Kindo}}, \bibinfo {author} {\bibfnamefont {L.}~\bibnamefont {Zhang}}, \bibinfo {author} {\bibfnamefont {E.}~\bibnamefont {Br{\"u}ck}}, \bibinfo {author} {\bibfnamefont {K.}~\bibnamefont {Buschow}}, \bibinfo {author} {\bibfnamefont {F.}~\bibnamefont {De~Boer}}, \bibinfo {author} {\bibfnamefont {C.}~\bibnamefont {Lef{\`e}vre}},\ and\ \bibinfo {author} {\bibfnamefont {G.}~\bibnamefont {Venturini}},\ }\bibfield  {title} {\bibinfo {title} {High-field magnetization of \text{RMn$_6$Sn$_6$} compounds with \text{R= Gd, Tb, Dy and Ho}},\ }\href {https://www.sciencedirect.com/science/article/pii/S0925838805005736} {\bibfield  {journal} {\bibinfo  {journal} {Journal of alloys and compounds}\ }\textbf {\bibinfo {volume} {408}},\ \bibinfo {pages} {169} (\bibinfo
  {year} {2006})}\BibitemShut {NoStop}%
\bibitem [{\citenamefont {Wang}\ \emph {et~al.}(2006)\citenamefont {Wang}, \citenamefont {Yates}, \citenamefont {Souza},\ and\ \citenamefont {Vanderbilt}}]{wang2006ab}%
  \BibitemOpen
  \bibfield  {author} {\bibinfo {author} {\bibfnamefont {X.}~\bibnamefont {Wang}}, \bibinfo {author} {\bibfnamefont {J.~R.}\ \bibnamefont {Yates}}, \bibinfo {author} {\bibfnamefont {I.}~\bibnamefont {Souza}},\ and\ \bibinfo {author} {\bibfnamefont {D.}~\bibnamefont {Vanderbilt}},\ }\bibfield  {title} {\bibinfo {title} {Ab initio calculation of the anomalous {H}all conductivity by {W}annier interpolation},\ }\href {https://journals.aps.org/prb/abstract/10.1103/PhysRevB.74.195118} {\bibfield  {journal} {\bibinfo  {journal} {Physical Review B}\ }\textbf {\bibinfo {volume} {74}},\ \bibinfo {pages} {195118} (\bibinfo {year} {2006})}\BibitemShut {NoStop}%
\bibitem [{\citenamefont {Ke}(2019)}]{ke2019intersublattice}%
  \BibitemOpen
  \bibfield  {author} {\bibinfo {author} {\bibfnamefont {L.}~\bibnamefont {Ke}},\ }\bibfield  {title} {\bibinfo {title} {Intersublattice magnetocrystalline anisotropy using a realistic tight-binding method based on maximally localized {W}annier functions},\ }\href {https://journals.aps.org/prb/abstract/10.1103/PhysRevB.99.054418} {\bibfield  {journal} {\bibinfo  {journal} {Physical Review B}\ }\textbf {\bibinfo {volume} {99}},\ \bibinfo {pages} {054418} (\bibinfo {year} {2019})}\BibitemShut {NoStop}%
\bibitem [{\citenamefont {Marzari}\ and\ \citenamefont {Vanderbilt}(1997)}]{marzari1997maximally}%
  \BibitemOpen
  \bibfield  {author} {\bibinfo {author} {\bibfnamefont {N.}~\bibnamefont {Marzari}}\ and\ \bibinfo {author} {\bibfnamefont {D.}~\bibnamefont {Vanderbilt}},\ }\bibfield  {title} {\bibinfo {title} {Maximally localized generalized {W}annier functions for composite energy bands},\ }\href {https://journals.aps.org/prb/abstract/10.1103/PhysRevB.56.12847} {\bibfield  {journal} {\bibinfo  {journal} {Physical review B}\ }\textbf {\bibinfo {volume} {56}},\ \bibinfo {pages} {12847} (\bibinfo {year} {1997})}\BibitemShut {NoStop}%
\bibitem [{\citenamefont {Souza}\ \emph {et~al.}(2001)\citenamefont {Souza}, \citenamefont {Marzari},\ and\ \citenamefont {Vanderbilt}}]{souza2001maximally}%
  \BibitemOpen
  \bibfield  {author} {\bibinfo {author} {\bibfnamefont {I.}~\bibnamefont {Souza}}, \bibinfo {author} {\bibfnamefont {N.}~\bibnamefont {Marzari}},\ and\ \bibinfo {author} {\bibfnamefont {D.}~\bibnamefont {Vanderbilt}},\ }\bibfield  {title} {\bibinfo {title} {Maximally localized {W}annier functions for entangled energy bands},\ }\href {https://journals.aps.org/prb/abstract/10.1103/PhysRevB.65.035109} {\bibfield  {journal} {\bibinfo  {journal} {Physical Review B}\ }\textbf {\bibinfo {volume} {65}},\ \bibinfo {pages} {035109} (\bibinfo {year} {2001})}\BibitemShut {NoStop}%
\bibitem [{\citenamefont {Marzari}\ \emph {et~al.}(2012)\citenamefont {Marzari}, \citenamefont {Mostofi}, \citenamefont {Yates}, \citenamefont {Souza},\ and\ \citenamefont {Vanderbilt}}]{marzari2012maximally}%
  \BibitemOpen
  \bibfield  {author} {\bibinfo {author} {\bibfnamefont {N.}~\bibnamefont {Marzari}}, \bibinfo {author} {\bibfnamefont {A.~A.}\ \bibnamefont {Mostofi}}, \bibinfo {author} {\bibfnamefont {J.~R.}\ \bibnamefont {Yates}}, \bibinfo {author} {\bibfnamefont {I.}~\bibnamefont {Souza}},\ and\ \bibinfo {author} {\bibfnamefont {D.}~\bibnamefont {Vanderbilt}},\ }\bibfield  {title} {\bibinfo {title} {Maximally localized {W}annier functions: Theory and applications},\ }\href {https://journals.aps.org/rmp/abstract/10.1103/RevModPhys.84.1419} {\bibfield  {journal} {\bibinfo  {journal} {Reviews of Modern Physics}\ }\textbf {\bibinfo {volume} {84}},\ \bibinfo {pages} {1419} (\bibinfo {year} {2012})}\BibitemShut {NoStop}%
\bibitem [{\citenamefont {Mostofi}\ \emph {et~al.}(2014)\citenamefont {Mostofi}, \citenamefont {Yates}, \citenamefont {Pizzi}, \citenamefont {Lee}, \citenamefont {Souza}, \citenamefont {Vanderbilt},\ and\ \citenamefont {Marzari}}]{mostofi2014updated}%
  \BibitemOpen
  \bibfield  {author} {\bibinfo {author} {\bibfnamefont {A.~A.}\ \bibnamefont {Mostofi}}, \bibinfo {author} {\bibfnamefont {J.~R.}\ \bibnamefont {Yates}}, \bibinfo {author} {\bibfnamefont {G.}~\bibnamefont {Pizzi}}, \bibinfo {author} {\bibfnamefont {Y.-S.}\ \bibnamefont {Lee}}, \bibinfo {author} {\bibfnamefont {I.}~\bibnamefont {Souza}}, \bibinfo {author} {\bibfnamefont {D.}~\bibnamefont {Vanderbilt}},\ and\ \bibinfo {author} {\bibfnamefont {N.}~\bibnamefont {Marzari}},\ }\bibfield  {title} {\bibinfo {title} {An updated version of wannier90: A tool for obtaining maximally-localised {W}annier functions},\ }\href {https://www.sciencedirect.com/science/article/pii/S001046551400157X} {\bibfield  {journal} {\bibinfo  {journal} {Computer Physics Communications}\ }\textbf {\bibinfo {volume} {185}},\ \bibinfo {pages} {2309} (\bibinfo {year} {2014})}\BibitemShut {NoStop}%
\end{thebibliography}
\end{document}